\documentstyle[prb,aps,multicol,epsf]{revtex}

\begin{document}

\title{Discrete aqueous solvent effects and possible attractive forces\footnote{Submitted to J. Chem. Phys.}}
\author{Y. Burak\footnote{email: yorambu@post.tau.ac.il} and D. Andelman\footnote{email: andelman@post.tau.ac.il}}
\address{School of Physics and Astronomy, Raymond and Beverly Sackler
  Faculty of Exact Sciences, \\ Tel Aviv University,
Tel Aviv 69 978, Israel}
\date{July 3, 2000}
\maketitle


\begin{abstract}
We study discrete solvent effects on the interaction of two
parallel charged surfaces in ionic
aqueous solution. These effects are
taken into account by adding a bilinear non-local term to the free
energy of Poisson-Boltzmann theory. We study numerically the
density profile of ions between the two plates, and the 
resulting inter-plate pressure.
At large plate separations the two plates are decoupled and the
ion distribution can be characterized by an effective
Poisson-Boltzmann charge that is smaller than the nominal charge.
The pressure is thus reduced relative to Poisson-Boltzmann
predictions. At plate
separations below $\sim 2\,{\rm nm}$ the pressure is modified
considerably, due to the solvent mediated short-range attraction
between ions in the the system. For high surface
charges this contribution can overcome the mean-field repulsion
giving rise to a net attraction between the plates.
\end{abstract}


\begin{multicols}{2}

\section{Introduction}
\label{sec:intro}

Aqueous ionic solutions are abundant in biological and chemical
systems. Often they play a prominent role in determining the
properties of charged macromolecules that are immersed in them
\cite{Israelachvili}. The mean field theory of electrolytes, known
as Poisson-Boltzmann (PB) theory and its linearized version,
Debye-H\"{u}ckel theory
\cite{Israelachvili,Gouy10,Chapman13,DH23,DH24,Andelman95}, are
known for many decades and have proved to be useful and important
tools. PB theory was applied in the study of colloidal dispersions
\cite{VO48,ACGMPH84}, biological membranes \cite{Andelman95},
synthetic and biological polyelectrolytes
\cite{Oosawa,BarratJoanny96}, and complex systems such as
DNA-lipid complexes\cite{HMGB98}. Nevertheless, PB theory is known
to have important limitations. Being a mean field theory, ion-ion
correlations are ignored. In addition, the finite size of ions is
neglected. These effects have been studied extensively using
various approaches \cite{BlumHenderson92} such as liquid
state
\cite{KjellanderMarcelja85,Kjellander88,LOB81,OuthBhu83}
and
density functional\cite{StevensRobbins90} theories, simulations
\cite{GJWL84,KAJM92,GKA97}, field theory
\cite{NetzOrland99EPL,NetzOrland00} and other modifications to the
PB theory \cite{BAO97,BAO00,LukSaf99,LZB99}.

Most of the studies of corrections to PB have concentrated on the
so-called primitive model, where ions are assumed to interact with
each other through the electrostatic interaction and a hard core
steric repulsion. Although this model can describe many effects
that are neglected in PB theory, it still neglects some
physical features that are present in real systems. Most
notably, the aqueous solvent is treated as a continuous medium,
whereas in reality ions interact with discrete solvent molecules.

Solvent effects are strong especially in water, because the
polar water molecules interact very strongly with ions. The most
significant result is that the electrostatic ion-ion interaction
is reduced by a factor $\varepsilon \simeq 78$ at room
temperature, due to screening by the dielectric environment.
However the discreteness of the solvent results in a more
complicated picture. When ions approach each other at separations
of a few water molecular diameters, the effective interaction
between them is modified considerably. Fig.~1 shows the correction
to the $1/\varepsilon r$ potential between two Na$^+$ ions in
water. This effective potential was calculated, using 
a simulation scheme
\cite{Lyubartsev97}, for a bulk NaCl solution of concentration
$0.55\,{\rm M}$, at room temperature.  Note that the short-range
potential, remaining after the subtraction of the Coulomb
interaction, is oscillatory and predominantly attractive.

The possibility to calculate the effective potential 
between ions in
water leads naturally to the model depicted schematically in
Fig.~2. The water is treated as a continuous medium, with a
dielectric constant $\varepsilon$. In addition to the
electrostatic interaction a short-range interaction is included
between ion pairs. The short-range potential, denoted as
$u_{ij}(r)$, is taken as an input to the model (from simulation), and
can in general depend on the ion species $i$ and $j$. For example,
the potential shown in Fig.~1 is used between Na$^+$-Na$^+$ pairs.
The effective potential is calculated in a bulk solution and thus
depends only on the ion-ion separation. However, systems containing
charged surfaces can lead to inhomogeneity or anisotropy in the
ion distribution.

The model described above was suggested in
Ref.~\onlinecite{Marcelja97Nature}, and was studied in planar
geometry using the Anisotropic Hyper-Netted Chain (AHNC)
approximation \cite{Kjellander88} in
Refs.~\onlinecite{Marcelja98,Marcelja97,KLM00,OttoPatey00}. In
Ref.~\onlinecite{BurakAndelman00a} we presented a simplified
approach to the same model. In this latter approach, a term
accounting for the short-range solvent-mediated ion-ion
interaction is added to the PB free energy. The formalism obtained
in this way is simple although less accurate than the AHNC
approximation, and in particular neglects ion-ion correlations. On
the other hand numerical calculations can be done fairly easily,
and are feasible in non-planar geometries. In addition, various
analytical results can be obtained, and the discrete solvent
effects can be readily understood in terms of basic physical
principles. In the present paper, which can be regarded as a
follow-up of Ref.~\onlinecite{BurakAndelman00a}, we use the same
formalism to study discrete solvent effects on interacting charged
and planar plates.

The outline of the paper is as follows. Section \ref{sec:model}
reviews the model and discusses its application to two charged and
planar plates. In Sec.~\ref{sec:density} we discuss the
corrections to the PB density profile. In section
\ref{sec:pressure} we obtain expressions for the 
inter-plate pressure and
derive a generalized contact theorem. The resulting pressure
curves are studied numerically and analytically in section
\ref{sec:curves}. Finally, section \ref{sec:conc} offers some
concluding remarks. The technical details in the derivation of the
pressure are presented in the Appendix.

\section{The Model}
\label{sec:model}

\subsection{Free energy}

The free energy of the system can be written
as a functional of the local ion densities,
consisting of the
usual PB term and a
hydration correction term. Assuming that the boundary
conditions are of fixed charges,
the following approximated form for the free energy
can be obtained \cite{BurakAndelman00a}:

\begin{eqnarray}
\Omega & = & \frac{\varepsilon}{8\pi} \int (\nabla \Psi)^{2}
   {\rm d}^{3}{\bf r} + k_{B}T
   \int \sum_{i} c_{i} \left( \ln
          \frac{c_{i}}{\zeta_{i}} - 1 \right)
   {\rm d}^{3}{\bf r}
\nonumber \\
 & & + \int \Lambda ({\bf r}) \left(
   {\bf \nabla}^{2}\Psi + \frac{4\pi}{\varepsilon}\sum_{i}c_{i}e_{i}
                              \right)
   \,{\rm d}^{3}{\bf r}
\nonumber \\
 & & + \frac{k_{B}T}{2} \sum_{i,j} \int c_{i}({\bf
         r})c_{j}({\bf r'})
              U_{ij}({\bf r}-{\bf r'})\,
              {\rm d}^{3}{\bf r}{\rm d}^{3}{\bf r'}
\label{eq:Omega}
\end{eqnarray}

\noindent where $\Psi$ is the electrostatic potential, $c_{i}$ are
the ion densities, $e_{i}$ are their respective charges,
$\varepsilon$ is the dielectric constant, $k_{B}T$ is the thermal
energy and $U_{ij}$ is defined below. The bulk ion densities
$c_{b,i}$ are determined by the fugacities 
$\zeta_i = \exp(\beta \mu_i)/\lambda_T^3$, 
where $\mu_i$ are the chemical
potentials, $\lambda_T$ is the de Broglie thermal wave length and
$\beta = 1/k_B T$. The simple PB relation $c_{b,i} = \zeta_i$ is
altered with the inclusion of hydration interactions, as will be
explained below. A detailed discussion of the various
approximations involved in Eq.~(\ref{eq:Omega}) is given in
Ref.~\onlinecite{BurakAndelman00a}. Here we shall briefly
discuss
each of the terms, and outline the way in which
Eq.~(\ref{eq:Omega}) is obtained.

The first three terms in Eq.~(\ref{eq:Omega}) form the usual PB
expression for the free energy. The first term is the
electrostatic free energy and the second term is the entropy of
the ions. The electrostatic potential $\Psi$ is a
functional of the ion densities $c_{i}$, and is determined by the
Poisson equation and the boundary conditions imposed by the
surface charges. Instead of writing this dependence explicitly in
the free energy, it is convenient to add a third term to $\Omega$,
containing a Lagrange multiplier $\Lambda({\bf r})$.

The fourth term in Eq.~(\ref{eq:Omega}) accounts for the hydration
interaction, and is quadratic in the ion densities.
The weighted potential $U_{ij}$ is defined as:

\begin{equation}
U_{ij} = 1 - {\rm e}^{-\beta u_{ij}(|{\bf r}-{\bf r'}|)}
\label{eq:ueff}
\end{equation}

\noindent where $u_{ij}$ is the nominal short-range hydration
interaction between ions of species $i$ and $j$. To obtain
Eq.~(\ref{eq:Omega}) we first treat the short-range interaction
$u_{ij}$ using a virial expansion of the grand canonical
potential, keeping terms up to the quadratic order. The
electrostatic interaction is then treated exactly as in PB theory,
using a mean field approximation for the electrostatic potential
$\Psi$. As an alternative approach Eq.~(\ref{eq:Omega}) can be
obtained from a field theory expansion of the grand partition
function \cite{Netz99pr}.

\subsection{Density equations}

The density profiles are obtained by minimizing the free energy
$\Omega$ with respect to the ion densities $c_i$. The third term
in Eq.~(\ref{eq:Omega}), containing the Lagrange multiplier
$\Lambda({\bf r})$ allows us to regard the densities $c_{i}({\bf
r})$ and the electrostatic potential $\Psi({\bf r})$ as
independent fields, and require that $\Omega$ has an extremum with
respect to the three fields $c_{i}$, $\Psi$ and $\Lambda$.
Requiring that $\Omega$ has an extremum with respect to $\Psi$
gives:

\begin{equation}
\Lambda = \frac{\varepsilon}{4\pi}\Psi \label{eq:lambda}
\end{equation}

\noindent and the extremum condition with respect to $c_{i}$ then
gives:

\begin{equation}
\ln \frac{c_{i}({\bf r})}{\zeta_{i}} + \sum_{j}\int
      c_{j}({\bf r'}) U_{ij}({\bf r}-{\bf r'})\,
      {\rm d}^{3}{\bf r'}
      + \beta e_{i} \Psi ({\bf r}) = 0
\label{eq:c3d}
\end{equation}

\noindent where relation (\ref{eq:lambda}) has been substituted to
express $\Lambda$ in terms of $\Psi$. This equation is
supplemented by the Poisson equation:

\begin{equation}
\nabla^{2}\Psi = -\frac{4\pi}{\varepsilon}\sum_{i}e_{i}c_{i}
\label{eq:Poisson}
\end{equation}

\noindent Since Eq.~(\ref{eq:c3d}) is an integral equation, the
$c_{i}$ cannot be written as a simple function of $\Psi$.
Therefore a single equation for $\Psi$, analogous to the PB
equation, cannot be obtained, and we are left with the two coupled
equations (\ref{eq:c3d}) and (\ref{eq:Poisson}). These equations
should be solved together to obtain the electrostatic potential
and density profiles. For $U \rightarrow 0$,
Eq.~(\ref{eq:c3d}) reduces to the Boltzmann equation
$c_{i}=\zeta_{i}\exp(-\beta e_{i} \Psi)$. In the bulk $\Psi = 0$,
leading to the relation $c_{b,i} = \zeta_i$. Combining these
relations with Eq.~(\ref{eq:Poisson}) reproduces the PB equation:

\begin{equation}
{\bf \nabla}^{2}\Psi = -\frac{4\pi}{\varepsilon}
                  \sum_{i}c_{b,i}e_{i}{\rm e}^{-\beta e_{i}\Psi}
                  \hspace{1cm}{\rm (PB)}
\label{eq:PB}
\end{equation}

Equations (\ref{eq:c3d}) and (\ref{eq:Poisson}) were solved for a
single charged and planar plate in Ref.~\onlinecite{BurakAndelman00a}.
The treatment of two parallel plates is very similar, and is
outlined below for completeness. The system is shown schematically
in Fig.~2b. The plate positions are designated by $z = 0$ and $z =
d$, using the convention that these are the coordinates of closest
approach of the ions to the plates (while the potentials
$u_{ij}(r)$ are measured from the {\it centers} of the ions). The
two plates are negatively charged, each one with a uniform surface
charge $\sigma$. No discreteness of surface charge is taken into
account in the present work. We assume an electrolyte of valency
$z_{+}$:$z_{-}$, {\it i.e.}, a solution of positive and negative
ions of charges $e_{\pm}=\pm z_{\pm}e$, where $e$ is the electron
charge.

In order to simplify the equations further, the interactions
between the different pairs of ion species can be taken to be
equal, {\it i.e.}, $U_{ij}({\bf r})=U_{++}({\bf r}) \equiv U({\bf
r})$ where $U_{++}$ is the weighted potential between the
(positive) counterions. The exact choice of $U_{+-}$ and $U_{--}$
is expected to be of only minor significance, as the co-ions are
repelled from the surface neighborhood and only the positive
counterions reach high densities there. From charge neutrality we
have $c_b \equiv c_{b,+} = (z_-/z_+) c_{b,-}$ and similarly
$\zeta \equiv \zeta_+ = (z_-/z_+) \zeta_-$, where the relation
between $c_b$ and $\zeta$ will be determined later.

Due to the one-dimensional symmetry imposed by the charged and
planar planes, the integration in Eq.\ (\ref{eq:c3d}) can be
performed over the $x-y$ plane to obtain:

\begin{equation}
\label{eq:c1d} c_{\pm}(z) =
     \zeta_{\pm} {\rm e}^{\mp \beta e z_{\pm} \Psi}\exp
      \left[-\int_{0}^{d} c(z')B(z-z')\,{\rm d}z'\right]
\end{equation}

\noindent where $c = c_{+}+c_{-}$ is the total ion density and
$B(z)$ is the effective interaction between two layers of ions,
expressed as an integral of $U({\bf r})$ in the plane of constant
$z$. Using cylindrical coordinates:

\begin{equation}
B(z)=2\pi \int_{0}^{\infty}\rho\,{\rm d}\rho\,
        U(\sqrt{z^{2}+\rho^{2}})
\label{eq:Bdefinition}
\end{equation}

\noindent The Poisson equation (\ref{eq:Poisson}) reads:

\begin{eqnarray}
\frac{{\rm d}^{2}\Psi}{{\rm d}z^{2}} & =
    & \frac{4\pi e}{\varepsilon} \zeta z_{+}
    \left({\rm e}^{\beta e z_{-} \Psi}-
          {\rm e}^{-\beta e z_{+}\Psi} \right)
\nonumber \\
    & & \ \ \times \exp \left[-\int_{0}^{d} c(z')B(z-z')\,{\rm d}z'\right]
\label{eq:Poisson1d}
\end{eqnarray}

\noindent Equations (\ref{eq:c1d}) and (\ref{eq:Poisson1d}) are
supplemented by the following boundary conditions:

\begin{equation}
\left.\frac{{\rm d}\Psi}{{\rm d}z}\right|_{z=0}=
        -\frac{4\pi}{\varepsilon} \sigma \ \ ; \ \
\left.\frac{{\rm d}\Psi}{{\rm d}z}\right|_{z=d/2}=0
\label{eq:boundary}
\end{equation}

\noindent since the problem with two plates of equal charge
at $z = 0$ and $z = d$ of equal charge
is symmetric about the mid-plane $z = d/2$.

Finally, the relation between $\zeta$ and the bulk density $c_b$
can be obtained from Eq.~(\ref{eq:c1d}). We imagine that the two
plates are immersed in a bath of electrolyte. In the region
outside the plates
an equation similar to Eq.~(\ref{eq:c1d}) holds, where the
integration inside the exponent is performed in the external region.
Far away from the plates, as $\Psi$ becomes zero, $c_{+}$
and $c_{-}$ assume their asymptotic constant, bulk values. The
integrand inside the exponential can be replaced by $-(1 +
z_{+}/z_{-})c_b B(z-z')$ leading to the result:

\begin{equation}
c_b = \zeta\exp\left[-\left(1+\frac{z_{+}}{z_{-}}\right)
B_t c_b \right]
\label{eq:cb}
\end{equation}

\noindent where:

\begin{equation}
B_t \equiv \int_{-\infty}^{\infty}{\rm d}z \, B(z) =
      \int {\rm d}^{3}{\bf r} \, U({\bf r})
\label{eq:Bt}
\end{equation}

\noindent is also equal to $2B_{2}$, the second virial
coefficient. The limit $B_{t}c_{b} \rightarrow 0$ is the limit in
which the short-range interaction becomes negligible in the bulk.
In this limit the relation between the bulk density and fugacity
of Eq.~(\ref{eq:cb}) tends to the ideal gas relation $c_{b} =
\zeta = \exp (\beta \mu)/\lambda_T^3$.

In the next section we will concentrate on a symmetric $1:1$
electrolyte, where the equations (\ref{eq:c1d}) and
(\ref{eq:Poisson1d}) take the form:

\begin{eqnarray}
c_{\pm}(z) & =
    & \zeta {\rm e}^{\mp \beta e \Psi}\exp
      \left[-\int_{0}^{d} c(z')B(z-z')\,{\rm d}z'\right]
\nonumber \\ \frac{{\rm d}^{2}\Psi}{{\rm d}z^{2}} & =
    & \frac{8\pi e}{\varepsilon} \zeta \sinh \left(\beta e \Psi \right)
\nonumber \\ 
    & &\ \ \times
         \exp \left[-\int_{0}^{d} c(z')B(z-z')\,{\rm d}z'\right]
\label{eq:symmetric}
\end{eqnarray}

\noindent and:

\begin{equation}
c_b = \zeta\exp\left(-2 B_t c_b \right)
\end{equation}

\subsection{Definitions and parameters}
\label{subsec:scales}

For the short-range
ion-ion potential $u({\bf r}-{\bf r'})$ we use the
effective potential between \mbox{${\rm Na}^{+}$ - ${\rm Na}^{+}$}
ion pairs, shown in Fig.~1. For ion-ion separations below $2.9
\mbox{\AA}$ a hard core interaction is assumed. Fig.~3 shows the
effective layer-layer interaction $B(z)$, as was derived from this
potential using Eq.~(\ref{eq:Bdefinition}). This effective
interaction is mostly attractive, as $B(z)$ is negative on most of
its range, and has a characteristic range of approximately
$7\,\mbox{\AA}$. The structure of $B(z)$ reflects the oscillatory
behavior of $u({\bf r})$.

It is useful to introduce the length scales characterizing
the PB density profiles \cite{Andelman95}. The {\it Gouy-Chapman
length}, defined as $b = \varepsilon k_B T/(2\pi e |\sigma|$),
characterizes the width of the diffusive counterion layer close to
a single charged plate with a surface charge density $\sigma$, in
the absence of added salt. The {\it Debye-H\"uckel screening
length}, $\lambda_{\rm D} = (8\pi c_b e^{2}/\varepsilon k_B
T)^{-1/2}$, equal to $19.6\,\mbox{\AA}$ for $c_b = 0.025\,{\rm M}$
at room temperature characterizes the decay of the screened
electrostatic interaction in a solution with added salt. The
strength of the electrostatic interaction can also be expressed
using the {\it Bjerrum length}, $l_{\rm B} = e^{2}/(\varepsilon
k_B T)$. This is the distance at which the electrostatic
interaction between two unit charges 
in a dielectric medium becomes equal to the thermal
energy. It is equal to about $7 \mbox{\AA}$ in water at room
temperature. In terms of the Bjerrum length $b = e/2\pi l_B
|\sigma|$ and $\lambda_D = (8\pi c_b l_B)^{-1/2}$.

The inclusion of the hydration interaction introduces additional
length scales in the system. For the interaction of Figs.~2 and
3, the range of the interaction $d_{\rm hyd}$ is approximately
$7\,\mbox{\AA}$, over twice the hard core diameter $d_{\rm hc} =
2.9\mbox{\AA}$. The strength of the hydration interaction is
characterized by the second virial coefficient $B_t/2$, with $B_t
\simeq -(7.9\mbox{\AA})^3$ as is calculated from
Eq.~(\ref{eq:Bt}).

\section {Density profiles}
\label{sec:density}

Equations (\ref{eq:c1d}) and (\ref{eq:Poisson1d})
are a set of three nonlinear
integrodifferential equations. We treat them numerically using an
iterative scheme, based on the assumption that the positive ion
density profile is dominated by the electrostatic interaction. We
start with the PB profile and calculate iteratively corrections to
this profile, as result from Eqs.~(\ref{eq:c1d}) and
(\ref{eq:Poisson1d}). For a 1:1 electrolyte we iteratively solve
the equation:

\begin{eqnarray}
\frac{{\rm d}^{2}\Psi^{(n)}}{{\rm d}z^{2}} & = &
     \frac{8\pi e}{\varepsilon} \zeta \sinh\left(\beta e \Psi^{(n)} \right)
\nonumber \\
   & & \ \ \times
         \exp\left[-\int_{0}^{d} c^{(n-1)}(z')B(z-z')\,{\rm d}z'\right]
\label{eq:iter}
\end{eqnarray}

\noindent where the superscript $n$ stands for the $n$th
iteration,

\begin{eqnarray}
c^{(n)}_{\pm}(z) & \equiv & \zeta{\rm e}^{\mp \beta e \Psi^{(n)}}
\nonumber \\
& & \ \ \times
\exp\left[-\int_{0}^{d} c^{(n-1)}(z')B(z-z')\,{\rm d}z'\right]
\end{eqnarray}

\noindent and the zeroth order densities $c^{(0)}_{\pm}$ are taken
as the density profiles generated by the PB equation
(\ref{eq:PB}). The boundary conditions (\ref{eq:boundary}) are
satisfied by the electrostatic potential in all iterations.
The solution converges after several iterations. It is
interesting to note that the first iteration captures most of the
effect. This observation can lead to various analytical results,
as shown in Ref.~\onlinecite{BurakAndelman00a} for a single plate.

In the following sections we will concentrate on the pressure
between the plates. First we discuss briefly the modification to
the PB density profiles. Let us begin by considering a large plate
separation $d$. In this case the results are similar to the
single-plate case, since $d$ is larger than all other length
scales in the system, and we present them for completeness.

Fig.~4 shows the density profile of the positively charged
counterions (solid line) between two charged plates, with $d =
50\,\mbox{\AA}$. Only one half of the system is shown, since the
profile is symmetric around the mid-plane. The surface charge,
$|\sigma| = 0.333\,{\rm C/m^{2}}$ corresponds to approximately
$48\,\mbox{\AA}^{2}$ per unit charge. This is a typical high surface
charge obtained for mica plates. It corresponds to a Gouy-Chapman
length $b = 1.06\,\mbox{\AA}$, at a temperature of $298\,$K, with
$\varepsilon=78$. The electrolyte bulk concentration is $0.025\,$M,
corresponding to a Debye-H\"uckel screening length $\lambda_{\rm
D}=19.58\,\mbox{\AA}$ . The density profile is compared to the
result of PB theory (dotted line).

The main effect is that the short-range attraction draws
additional counterions to the vicinity of the charged plate. 
Note, however, that the contact density remains very close
to the PB density, as will be explained later. The
increase of the counterion density near the plate is followed by a
depletion further away. This can be understood in the no-salt case
since the total number of counterions is fixed. In our case the
salt concentration is low. The Debye-H\"{u}ckel screening length
is large compared to the Gouy-Chapman length and compared to the
range of the short-range interaction, so the salt has a minor
effect.

The counterion density profile is also compared with results of
the AHNC approximation \cite{Marcelja00pr} that were obtained
using the same short-range hydration potential (`x' symbols). The
qualitative effect is similar in our model and in the AHNC.
Specifically, both density profiles follow the PB density curve
for the first few Angstr\"oms 
from the plate and show a considerable decrease in
the positive ion density, relative to PB, starting at a distance
of about $5\,\mbox{\AA}$ from the plate. The maximal decrease in the
density is approximately
$30\%$ in our model and almost $50\%$ in the AHNC profile, both
relative to the PB profile.

The effect of the short-range ion-ion interaction strongly
depends on the surface charge. This is demonstrated in Fig.~5.
The ratio of the counterion density and its PB value,
$c/c^{\rm PB}$, is shown for
three values of $\sigma$. The effect of the hydration potential is
very minor for small surface charge ($|\sigma| = 0.0333\,{\rm
C/m^{2}}\simeq 1\,e/480\,\mbox{\AA}^2$), where 
the ratio $c/c^{\rm PB}$ is
approximately $2\%$ at its maximum, and considerable for a
surface charge of
$0.333\,{\rm C/m^{2}}=1\,e/48\,\mbox{\AA}^2$,
where it reaches approximately $40\%$.

As the plate separation decreases, the modification to
$c^{\rm PB}$ is
expected to remain similar to the single plate case as long as
$d/2$ is large compared to $b$ and to $d_{\rm hyd}$. This can
indeed be seen in Fig.~6, where a high surface charge, as in
Fig.~4, is considered. In this case $b \ll d_{\rm hyd}$, so a
deviation from the single plate curve is expected when $d/2
\lesssim d_{\rm hyd} \simeq 7\,\mbox{\AA}$. The ratio
$c/c_{\rm PB}$ is shown for several plate
separations between $5$ and $50\,\mbox{\AA}$. The results are very
similar for $d = 50, 35$ and $20\,\mbox{\AA}$ (Fig.~6a). In
particular, note that the contact density remains very close to
the PB value in all three separations. This is a result of the
generalized contact theorem, derived in section
\ref{sec:pressure}. For smaller $d$, $5$ and $10\,\mbox{\AA}$
(Figures 6b and 6c, respectively) the behavior is different, and
in particular the contact density deviates from the PB value. The
effect of decreasing $d$ was found to be
similar for smaller surface charge
({\it e.g.}, $0.1\,{\rm C/m^2}$,) and for salt concentration up to
$0.1\,{\rm M}$.

The most important effect on the density profile is that the ion
density is depleted far away from the charged plates. When the two
plates are highly separated from each other the ion density can be
described, far away from the plates, using an effective PB surface
charge. This effective charge was calculated in
Ref.~\onlinecite{BurakAndelman00a}, and is smaller than the
nominal charge (for example, for the surface charge used in Fig.~4
it is smaller by a factor of $\sim3.8$). The reduced density leads
to a reduced pressure, relative to PB, as will be explained in the
following sections.

\section {Pressure equation and contact theorem}
\label{sec:pressure}

The pressure $P_{\rm in}$ in the region between the two plates can
be obtained by differentiating the free energy $\Omega$ with
respect to the plate separation $d$:

\begin{equation}
P_{\rm in} = -\frac{\delta \Omega}{\delta d}
\end{equation}

\noindent To compute $\delta \Omega$ we can imagine that a `slice'
of width $\delta d$ is inserted at some position $z_{0}$
between the two plates. Adding
up all the contributions to $\delta \Omega$, and using Eq.\
(\ref{eq:c1d}) and the boundary conditions (\ref{eq:boundary}) we
obtain:

\begin{eqnarray}
\label{eq:pressurein} P_{\rm in} & = & k_{B}T\sum_{i}c_{i}(z_{0})
    - \left.\frac{\varepsilon}{8\pi}\left(\frac{{\rm d}\Psi}
      {{\rm d}z}\right)^2 \right|_{z_{0}}
\nonumber \\
& - & k_{B}T\sum_{ij}\int_{0}^{z_{0}}{\rm d}z
                   \int_{z_{0}}^{d}{\rm d}z' \,
                   c_{i}(z)c_{j}(z')
                   \frac{{\rm d}B}{{\rm d}z}(z'-z)
\end{eqnarray}

\noindent This result is correct for any combination of
ion species $i$, assuming the same short-range interaction
$u_{ij}$ between different ion pairs.
The full derivation is given in the Appendix.
The pressure is equal throughout the plate spacing and,
therefore, independent on the choice of $z_{0}$.

The net pressure $P$ between the plates is the difference
between the pressure inside and outside the plates. The latter is
equal throughout the region outside the plates. In particular, it
is equal to the bulk pressure $P_{\rm bulk}$, so we have:

\begin{equation}
P = P_{\rm in} - P_{\rm bulk}
\end{equation}

\noindent To obtain $P_{\rm bulk}$, we note that an
equation
similar to Eq.~(\ref{eq:pressurein}) holds
in the bulk, with constant
electrostatic potential and with $c_{i}$ constant and being
equal to the
bulk densities. For the case of a 1:1 electrolyte, we find:

\begin{equation}
\label{eq:pbulk} P_{\rm bulk} = 2k_B Tc_b \left(1+B_t
c_b\right)
\end{equation}

\noindent Since $B_t$ is negative the bulk pressure is lower than
its PB value.
Note that in the case of no added salt $P_{\rm bulk} = 0$.

The expression (\ref{eq:pressurein}) assumes a particularly simple
form if we set $z_{0}$ to zero, namely on one of the plates.
Then the third term in (\ref{eq:pressurein}) vanishes and the
second term is fixed by the boundary conditions, giving:

\begin{equation}
P = k_B T\sum_{i}c_{i}(0) - \frac{2\pi}{\varepsilon}\sigma^{2}
  - P_{\rm bulk}
\label{eq:pressure0}
\end{equation}

\noindent Alternatively, if we choose $z_{0}$ at the mid-plane,
$z = d/2$, by symmetry the second term in (\ref{eq:pressurein})
vanishes and the pressure is expressed as:

\begin{eqnarray}
P  & = & k_{B}T\sum_{i}c_{i}(d/2) \nonumber \\ & - &
k_{B}T\sum_{ij}\int_{0}^{d/2}{\rm d}z
                   \int_{d/2}^{d}{\rm d}z'
                   \, c_{i}(z)c_{j}(z')
                   \frac{{\rm d}B}{{\rm d}z}(z'-z)
\nonumber \\
    & - & P_{\rm bulk}
\label{eq:pressured}
\end{eqnarray}

\noindent The equality of these two expressions for the pressure
results in the generalized contact theorem
\cite{Israelachvili,CarnieChan81}:

\begin{eqnarray}
& & \sum_{i}  c_{i}(0) = \frac{2\pi \beta}{\varepsilon}\sigma^{2}
           + \sum_{i} c_{i}(d/2) \nonumber \\
& & - \sum_{ij}\int_{0}^{d/2}{\rm d}z \int_{d/2}^{d}{\rm d}z'
                     \, c_{i}(z)c_{j}(z')
                     \frac{{\rm d}B}{{\rm d}z}(z'-z)
\label{eq:contact}
\end{eqnarray}

\noindent The very small relative change of the contact density,
compared to PB theory, at large plate separations can be
understood from this result. We consider first the case of high
surface charge, where the the Gouy-Chapman length is small
compared to the Debye-H\"uckel screening length,
$b \ll \lambda_D$. In
this case the second and third terms on the right hand side of
Eq.~(\ref{eq:contact}) become negligible compared to the first
term when $d \gg b, d_{\rm hyd}$, where $d_{\rm hyd}$ is the range
of the hydration interaction. The contact ion
density is then dominated by the positive ion density, and is very
close to the PB value. When there are only counterions in the
solution and $d \rightarrow \infty$ (or equivalently, in the case
of one isolated plate), we have exactly, as in PB theory :

\begin{equation}
c_{+}(0) = \frac{2\pi \beta}{\varepsilon}\sigma^{2} \ \ \ \ \
\mbox{(one plate, no salt)}
\end{equation}

If $b$ is not small compared to $\lambda_D$, the correction to the
contact density is still small for large enough plate separations,
assuming that the hydration interaction is negligible in the bulk,
{\it i.e.} $-B_t c_b = -B_t /(8\pi l_{\rm B}\lambda_{\rm D}^{2})
\ll 1$. When $d \gg \lambda_D$ and $d > d_{\rm hyd}$, 
the coupling between the two plates is negligible
and Eq.~(\ref{eq:contact}) becomes:

\begin{equation}
\sum_{i} c_{i}(0) \simeq \frac{2\pi \beta}{\varepsilon}\sigma^{2}
    + P_{\rm bulk} 
\label{eq:contact_ap}
\end{equation}

\noindent The only difference in this expression relative to the
PB contact density is the change in the bulk pressure. This change
is negligible if the hydration interaction is negligible in the
bulk.

For smaller $d$, the integral in (\ref{eq:contact}) can contribute
to a significant change in the contact density relative to PB
theory. This can be seen in Fig.~6, where $b << \lambda_{\rm D}$,
at plate separations below $\sim 10\,\mbox{\AA}$.


\section{Pressure curves}
\label{sec:curves}

\subsection{Pressure beyond Poisson-Boltzmann}
\label{subsec:beyondPB}

Using equations~(\ref{eq:pressured}) and (\ref{eq:pbulk}) the
pressure can be written in our model as the sum of the following
three terms:

\begin{eqnarray}
P & = & k_B T\sum_{i}\left[c_{i}(d/2)-c_{b,i}\right] \nonumber \\
   & - & k_{B}T\sum_{ij}\int_{0}^{d/2}{\rm d}z
                   \int_{d/2}^{d}{\rm d}z'
                   \, c_{i}(z)c_{j}(z')
                   \frac{{\rm d}B}{{\rm d}z}(z'-z)
                   \nonumber \\
   & - & 2k_B TB_t c_b^2
\label{eq:pressure1}
\end{eqnarray}

\noindent A symmetric 1:1 electrolyte is assumed
for simplicity throughout
this section. We would like to compare this pressure with the PB
pressure, which can be written as follows:

\begin{equation}
P_{\rm PB} = k_B T\sum_{i}\left[c_{{\rm PB},i}(d/2)-c_{b,i}\right]
\label{eq:ppb}
\end{equation}

\noindent where $c_{{\rm PB},i}(d/2)$ is the PB density of the
$i$th ion species at the mid-plane. The first term in
Eq.~(\ref{eq:pressure1}),

\begin{equation}
P_m = k_B T\sum_{i}\left[c_{i}(d/2)-c_{b,i}\right]
\label{eq:pm}
\end{equation}

\noindent is similar in form to the PB pressure
(\ref{eq:ppb}), but the mid-plane
density in equations (\ref{eq:pm}) and (\ref{eq:ppb}) can be
different. The second term in Eq.~(\ref{eq:pressure1}), which we
denote as the hydration pressure:

\begin{equation}
P_{\rm hyd} = - k_{B}T\sum_{ij}\int_{0}^{d/2}{\rm d}z
                   \int_{d/2}^{d}{\rm d}z'
                   \, c_{i}(z)c_{j}(z')
                   \frac{{\rm d}B}{{\rm d}z}(z'-z)
\label{eq:phyd}
\end{equation}

\noindent is the integrated short-range force acting
between ion pairs in the two halves of the system. The third term
is the change in the bulk pressure
relative to PB theory, due
to the inclusion of a 2nd virial coefficient in the bulk equation
of state.

Some simple observations can be made immediately from
Eq.~(\ref{eq:pressure1}). These observations will be useful in the
next subsection, where the numerically calculated pressure curves
are presented (Figs.~7 and 8). For now let us assume that the
third term in Eq.~(\ref{eq:pressure1}) is negligible as compared to
the first two. Of these two terms, the first, $P_m$, is linear
in the density whereas the second term, $P_{\rm hyd}$, is
quadratic. As a result, the relative importance of $P_m$ and
$P_{\rm hyd}$ depends on the plate separation $d$. At large $d$
the density in the mid-plane region is small, so that $P_{\rm hyd}
\ll P_m$. The main correction to to the PB pressure (\ref{eq:ppb})
then comes from the change of the mid-plane density, $c(d/2)-c_{\rm
PB}(d/2)$. Far away from the two plates the system behaves as
predicted by PB theory with a modified, effective surface charge.
The mid-plane density is depleted relative to PB, since
counterions are attracted to the vicinity of the charged plates.
Hence the pressure is smaller than in PB theory. As the plate
separation decreases and the mid-plane density increases, $P_{\rm
hyd}$ can become important.

\subsection{Numerical results}

The general arguments of the previous section can be verified by
calculating numerically the pressure using
Eq.~(\ref{eq:pressure1}). Fig.~7 shows the pressure as a function
of the plate separation $d$ for a surface charge $|\sigma| =
0.333\,{\rm C/m^{2}} \simeq 1\,e/48\,\mbox{\AA}^2$ and bulk ion
density $c_b = 0.025$M (solid line). 
The pressure is compared with
$P_{\rm PB}$ (dotted line). The contribution of $P_m$, the first
term in Eq.~(\ref{eq:pressure1}), is also shown (dashed line).

The behavior of the pressure at a large range of plate separations
is shown in Fig.~7a on a semi-logarithmic plot. At large $d$, the
pressure is dominated by $P_m$, as expected. It is considerably
smaller than the PB pressure, due to the reduced effective charge
on the plates. At lower $d$ the second term in
Eq.~(\ref{eq:pressure1}), $P_{\rm hyd}$, becomes dominant,
and the
overall interaction is attractive at plate separations between $6$
and $12\,\mbox{\AA}$. Note that the apparent sharp decrease in the
pressure at a separation of approximately $13\mbox{\AA}$ is
artificial, and results from the divergence of the logarithmic
scale as the pressure approaches zero. Fig.~7b shows the 
same pressure
using a linear scale, in the region in which it becomes negative
(attractive). The net pressure crosses 
smoothly from positive to negative
values due to a steady increase in the magnitude of the (negative)
$P_{\rm hyd}$. At very short separations $P_m$ dominates again,
and the pressure coincides with the predictions of PB theory.

Fig.~8 shows the effect of the hydration potential for a smaller
surface charge, $|\sigma| = 0.119{\rm C/m^{2}} \simeq 1\,e/
135\,\mbox{\AA}^2$.
In this case and for all surface charge, 
$|\sigma| \lesssim 0.25\,{\rm C/m^2}$, 
the pressure is repulsive at all plate separations.
The correction over the PB result is much
smaller than in Fig.~7, but still significant. At plate separations of
approximately $5$ to $20 \mbox{\AA}$ $P_{\rm hyd}$ is the dominant
contribution to the deviation from PB, and results in a
considerably reduced pressure. At larger $d$ the pressure is
reduced mainly because of the change in the mid-plane density.

\subsection{Comparison with AHNC}

Fig.~9 shows a comparison of the pressure obtained 
in our model (a) and in
the AHNC approximation\cite{Footnote1} (b), at surface
charge $|\sigma| = 0.333\,{\rm C/m^{2}} \simeq
1\,e/48\,\mbox{\AA}^2$. The same short-range hydration potential
is used in the two calculations. The main plots show the pressure
using a logarithmic scale. The insets show the pressure on a
linear scale in the region where it becomes attractive. The full
pressure (solid line) is compared in Fig.~9a with the PB pressure
(dotted line). In Fig~9b the AHNC pressure (solid line) is
compared with the pressure obtained using an electrostatic and
hard core interaction only (dotted line). Since the AHNC
approximation accounts for ion-ion correlations, there are
differences between the pressure curves in our model
as compared to the AHNC approximation.
However a comparison of Figs.~9a and 9b
shows that very similar qualitative and even semi-quantitative
effects of the hydration interaction are found in the two
calculations.

A comparison for smaller $|\sigma| = 0.119{\rm C/m^{2}} \simeq
1\,e/ 135\,\mbox{\AA}^2$ is shown in Fig.~10. The solid line is the
pressure in our model and the dashed line is the AHNC pressure.
The dotted line shows the PB pressure. As in Fig.~9, the
qualitative effect is similar in the two calculations.

Since the AHNC approximation takes into account ion-ion
correlations, the comparison allows us to assess the relative
importance of correlations and discrete solvent effects. The
results shown in Figs.~7 and 9 indicate that discrete solvent effects
can be much larger than correlation effects induced by the
electrostatic interaction. For smaller surface charge, as in
Figs.~8 and 10, these effects are of similar order of magnitude.
In the AHNC approximation the pressure includes an electrostatic
term due to correlations between ions in the two halves of the
system, in addition to the hydration and mid-plane density
contributions. In Fig.~10 this term is of similar order of
magnitude as $P_{\rm hyd}$, and is the main source for the
difference between the solid line (our model) and dashed line
(AHNC). For larger surface charge, as in Fig.~9, $P_{\rm hyd}$
becomes much larger than the electrostatic contribution.

When divalent ions are present in the solution, correlation
effects become much larger than in the monovalent case
\cite{GJWL84,KAJM92}. Discrete
solvent effects are also modified, since the effective short-range
interaction between two divalent ions is different in the two
cases. When the charge
on the ions is doubled, the electrostatic interaction between two
ions increases by a factor of $4$. The ion-ion separation where
the electrostatic interaction is equal to $k_B T$ increases from
$l_B \simeq 7\,\mbox{\AA}$ to almost $30\,\mbox{\AA}$. On this
electrostatic scale, the water molecular size ($\sim
3\,\mbox{\AA}$) is much smaller than in the monovalent case.
Hence we can expect the solvent to be
more similar to a continuous dielectric medium.
Indeed, the correction to the $1/\varepsilon r$ potential between
two (artificial)
Na$^{2+}$ ions in water \cite{GuaPad96} is found to be purely
repulsive, and is significant only at separations below $\sim
10\,\mbox{\AA}$, where the electrostatic interaction is
considerably larger than $k_B T$. Thus we expect discrete solvent
effects to be less important than correlations in the
divalent case.

\subsection{Further analysis}
\label{subsec:further}

\subsubsection{Large plate separations}

As discussed above, the hydration term becomes small at large
$d$, compared to the change in the mid-plane
density. In order to study the contribution of the mid-plane
density to the pressure, let us assume that the plate separation
$d$ is much larger than all other length scales in the system:
$b, d_{\rm hyd}, \lambda_D$.
The two plates are then decoupled and the
mid-plane potential can be written as $\Psi (d/2) \simeq 2
\Psi_1(d/2)$ where $\Psi_1(d/2)$ is the electrostatic potential at
a distance $d/2$ from a {\it single} plate. We assume also that
$\lambda_D \gg b$, which is usually the case
when the surface charge density is large. At a large distance from
the plate the single plate profile is a PB profile, corresponding
to a renormalized surface charge $\sigma_{\rm eff}$
\cite{BurakAndelman00a}. The contribution $P_m$ to the pressure
can then be written as follows \cite{Andelman95}:

\begin{equation}
P_{\rm m} \simeq \frac{8k_B T}{\pi l_B \lambda_D^2}
            \left(1-\frac{2b_{\rm eff}}{\lambda_D}\right)
            {\rm e}^{-d/\lambda_D}
\end{equation}

\noindent where $b_{\rm eff} = 1/2\pi l_B |\sigma_{\rm eff}|$ is
the effective Gouy-Chapman length. A similar expression holds for
the PB pressure, with the nominal Gouy-Chapman length $b$ used
instead of $b_{\rm eff}$. We thus find that:

\begin{equation}
\frac{P_m}{P_{\rm PB}} \simeq
      \frac{1 - 2b_{\rm eff}/\lambda_D}{1 - 2b/\lambda_D}
      \simeq 1 - 2\frac{b_{\rm eff} - b}{\lambda_D}
\end{equation}

\noindent In Ref.~\onlinecite{BurakAndelman00a} an analytical
expression for $b_{\rm eff} - b$ is found. Its general behavior is:

\begin{equation}
b_{\rm eff} - b \sim -\frac{B_t}{l_B b}
\end{equation}

\noindent with a numerical prefactor of $1/12\pi$ in the limit
$b \ll d_{hyd}$
and a numerical prefactor $1/4\pi$ in the limit $b \gg d_{hyd}$.
The parameters of the hydration interaction
$d_{hyd} \simeq 7\,\mbox{\AA}$
and $B_t \simeq -500\,\mbox{\AA}^3$ are as defined in
Sec.~\ref{subsec:scales}.

A careful treatment of Eq.~(\ref{eq:pressure1}) shows that the
second and third terms also add a contribution to the pressure
that should be regarded as linear in the density, although this
contribution is small. For large enough $d$ the
integration range in the second term of Eq.~(\ref{eq:pressure1})
can be extended to be between $-\infty$ and $+\infty$ because
${\rm d}B/{\rm d}z$ has a finite range. In addition all quantities
can be replaced by their mid-plane values. We then find that the
second and third terms of Eq.~(\ref{eq:pressure1}) give:

\begin{eqnarray}
   & - & k_B T\sum_{ij}\int_{0}^{d/2}{\rm d}z
                   \int_{d/2}^{d}{\rm d}z'
                   \, c_{i}(z)c_{j}(z')
                   \frac{{\rm d}B}{{\rm d}z}(z'-z)
\nonumber \\
   & - & 2k_B TB_t c_b^2
\nonumber \\
   & \simeq & \frac{1}{2}k_B T B_t \sum_{ij}[c_i(d/2)-c_b][c_j(d/2)-c_b]
\nonumber \\
   & & + 2 k_B T B_t c_b \sum_{i}[c_i(d/2) - c_b]
\end{eqnarray}

\noindent The first term is quadratic in $[c_i(d/2) - c_b]$ and
can be neglected relative to $P_m$ at large $d$. The second term
is linear, although small because $B_t c_b \ll 1$. It accounts for
the small difference between the dashed and solid lines at large
$d$ in Fig.~7a.

\subsubsection{Hydration pressure}

The behavior of $P_{\rm hyd}$, the hydration pressure term, can be
understood as follows. As a zero-th order approximation, the ion
density is dominated by electrostatics and can be replaced in
Eq.~(\ref{eq:phyd}) by its PB value. Fig.~11 shows that this gives
a very good approximation. Hence we can write:

\begin{equation}
P_{\rm hyd} \simeq \sum_{ij}\int_{0}^{d/2}{\rm d}z
                   \int_{d/2}^{d}{\rm d}z'
                   \, c_{{\rm PB},i}(z)
                   F(z'-z)c_{{\rm PB},j}(z')
\label{eq:phydF}
\end{equation}

\noindent where

\begin{equation}
F(z) \equiv -k_B T\frac{{\rm d}B(z)}{{\rm d}z}
\end{equation}

\noindent represents the force between two planar ion layers
separated by a distance $z$. The following behavior of $F(z)$ can
be inferred from Fig.~3. At inter-layer separations  $z < d_{\rm
hc} = 2.9\,\mbox{\AA}$ $F(z)$ is positive (repulsive). At larger
$z$ the value of $B(z)$ increases from its large negative value at
$z = d_{\rm hc}$ to zero over a few Angstr\"{o}ms, leading to a
strongly attractive (negative) $F(z)$. A closer inspection of
Fig.~3 shows that $F(z)$ is oscillatory, due to the local maxima
and minima of $B(z)$. As we shall see below these fine details
are smoothed away when two diffusive layers of finite thickness
interact.

The behavior of $P_{\rm hyd}$ in Fig.~11 can now be understood as
follows. Most of the counterions are concentrated near the two
plates, in layers whose thickness is of order $b =
1.06\,\mbox{\AA}$. Note that $b$ is small compared to $d_{\rm hyd}
\simeq 7\,\mbox{\AA}$. When $d > d_{\rm hyd}$ these two layers do
not interact directly with each other through the short-range
interaction. Ions in the two sides of the mid-plane interact with
each other, leading to a negative (attractive) $P_{\rm hyd}$. As
$d$ is decreased towards $d_{\rm hyd}$, larger and larger ion
densities come into contact through $F(z)$ and the magnitude of
the negative $P_{\rm hyd}$ increases accordingly. The gradual
increase in the magnitude of $P_{\rm hyd}$ reflects the algebraic
decay of the density profile near each layer. When $d$ decreases
below $\sim 2d_{\rm hyd} \simeq 14\,\mbox{\AA}$, the magnitude
of $P_{\rm hyd}$ increases more rapidly, as the ions in the
two layers interact with ions in the mid-plane region.

The behavior of $P_{\rm hyd}$ changes when $d$ decreases below
$d_{\rm hyd}$. Most of the contribution to $P_{\rm hyd}$ now comes
from the interaction between the dense counterion layers near the
two plates. As $d$ decreases these layers are separated by
correspondingly decreasing distances. The hydration pressure
follows roughly the structure of $F(z)$. It is strongly attractive
for $d \gtrsim d_{\rm hc}$ and repulsive for $d < d_{\rm hc}$. The
fine details of $F(z)$ are smoothed due to the thickness of the
diffusive ion layers.

As the plate separation decreases below $d_{\rm hc}$ towards
contact $P_{\rm hyd}$ tends to zero, as it should since 
$F(0) = -k_B T\left.\frac{{\rm d}B}{{\rm d}z}\right|_{z=0} = 0$.
One implication of this result
is that $P_m$ returns to be the dominant contribution to the
pressure, even for high surface charges. Another implication is
that the short-range interaction becomes unimportant. As in PB
theory, the ions in the region between the two plates become
essentially a confined ideal gas, and their total number is
determined by charge neutrality. Thus $P_m$ coincides with the PB
pressure matching the {\it nominal} surface charge density
$\sigma$. This is seen clearly in Fig.~7.

\subsubsection{Small plate separations}

In experiments the actual surface charge is usually not
exactly known,
because the number of ions dissociating from the surface is
uncontrolled. The PB charge is then fitted to the large separation
behavior. This charge can be significantly smaller than the actual
surface charge, as discussed above. The interpretation of our
results is then as follows. At plate separations below
approximately $2$ nm, an attractive force appears, due to $P_{\rm
hyd}$. This force can reduce the net repulsion, or even induce a
net attraction, depending on the surface charge on the plates. As
the plate separation decreases below the range of the hydration
interaction $d_{\rm hyd} \simeq 7\,\mbox{\AA}$, $P_{\rm hyd}$
decreases and eventually tends to zero. The pressure then
coincides with the PB pressure matching the {\it nominal} surface
charge. As was pointed out in Ref.~\onlinecite{Marcelja97Nature} this
leads to an apparent strong repulsive force when compared with the
PB curve fitted to the large separation behavior. As an example,
the pressure corresponding to $\sigma = 0.25\,{\rm C/m}^2 \simeq
1\,e/64\,\mbox{\AA}^2$ is shown in Fig.~12 as a function of
$d$ (solid line) using a linear scale. The dashed line shows the
PB pressure curve using an effective surface charge chosen to
match the large $d$ behavior of the solid line. When the two lines
are compared a strong (apparent) repulsive contribution is seen in
the solid line below $d \simeq 5\,\mbox{\AA}$, and an attractive
contribution is seen for $5\,\mbox{\AA} \lesssim d \lesssim
15\,\mbox{\AA}$.

\section{Concluding remarks}
\label{sec:conc}

Summarizing our results on the pressure, we find that hydration
effects can be understood as arising from two contributions. The
first contribution is the change in the mid-plane ion density.
This contribution dominates at large plate separations and can be
understood in terms of an effective PB surface charge in our
model. The effective PB charge is smaller than the nominal charge
due to the accumulation of counterions in the vicinity of the
charged plates.
Thus the pressure is reduced relative to PB theory, using
the same surface charge. 

As an alternative viewpoint, the PB surface charge can be
chosen to match the large plate separation of the pressure
in our model. When this is done, 
an apparent repulsive force appears in our model at 
very small plate separations ($\lesssim 5\,\mbox{\AA}$),
as compared with the fitted PB pressure.

The second contribution to the pressure is the direct solvent
mediated attraction between ion pairs in the two halves of the
system. This latter term can become dominant at plate separations
between $\sim$$0.5\,{\rm nm}$ and $\sim$$2\,{\rm nm}$. It can
induce a net attractive interaction between the two plates when
the surface charge is high.

Attraction between like-charged surfaces is never predicted by PB
theory \cite{Neu99,SaderChan99}. On the other hand, mechanisms
involving correlations can lead to attraction. Several approaches
have shown that ion-ion correlations can have this effect, in the
framework of the primitive model
\cite{StevensRobbins90,GJWL84,KAJM92}. In practice, this
attraction can be strong enough to overcome the mean field
repulsion when divalent ions are present in the solution. When
there are only monovalent ions in the solution, ion-ion
correlations have a much smaller effect. Another mechanism that
can lead to attraction is the van der Waals force, arising from
correlations between the polarizations on the two surfaces. As we
find in this work, solvent mediated forces, related to ion-solvent
correlations, are another mechanism that can induce inter-surface
attraction. In some cases (monovalent ions, small separation,
large surface charge) they are the leading mechanism for
attraction.


A strong deviation from PB predictions is indeed 
measured\cite{Pashley81a,PashIsrael84} between
charged surfaces in aqueous solution at
separation below $\sim 2 \,{\rm nm}$. 
The force includes an oscillatory
contribution, with a period corresponding to the water molecular
size. This force is due to the structuring of water in layers
between the surfaces. In addition to this oscillatory
contribution, an additional strong contribution is seen, which is
often referred to as the hydration force
\cite{Israelachvili,Pashley81a}.  The aqueous pair potential model
of Ref.~\onlinecite{Marcelja97Nature} was a first step towards the
understanding of this force. A more realistic picture will
probably emerge if a proper effective ion-surface interaction will
be included, in addition to the effective ion-ion interaction. In
addition, the modification to the ion-ion effective potential in a
confined geometry may also be important. In order to assess the
importance of these effects, further simulation results are needed
as an input to the model.

%
The aqueous pair potential and the free energy (\ref{eq:Omega})
involve various approximations, which are discussed extensively in
Refs.~\onlinecite{BurakAndelman00a,Marcelja99}. Nevertheless,
the large modification to the PB
pressure, as obtained also using the AHNC
approximation\cite{KLM00,OttoPatey00,Marcelja00pr}
, indicates that the solvent
effects on the ion distribution are a crucial ingredient in the
origin of hydration forces\cite{Marcelja97Nature,IsraelWenner96}.
The semi-quantitative agreement of our results with the AHNC
approximation indicates that our formalism captures the important
effects and suggests its further application in non-planar
geometries, where the AHNC approximation is not applicable.

\acknowledgements

We wish to thank S. Mar\u{c}elja for numerous valuable
discussions and for sharing with us his results prior to
publication.
Partial support from the U.S.-Israel Binational
Foundation (B.S.F.) under Grant No. 98-00429, and the Israel
Science Foundation founded by the Israel Academy of Sciences and
Humanities - Centers of Excellence Program is gratefully
acknowledged.

\end{multicols}

\appendix

\section{Derivation of the pressure}
\label{ap:pressure}

The free energy of the system is given by the
sum: $\Omega = \Omega_{\rm PB} +
\Delta \Omega$ with $\Omega_{\rm PB}$ and $\Delta \Omega$ defined
as follows:

\begin{eqnarray}
\Omega_{\rm PB} & = & \frac{\varepsilon}{8\pi}\int_{0}^{d}
         \left(\frac{{\rm d}\Psi}{{\rm d}z}\right)^{2}
         \, {\rm d}z +
        k_B T\int_{0}^{d} \sum_{i} c_{i}\left(
          \ln \frac{c_{i}}{\zeta}-1 \right)\,{\rm d}z
\nonumber \\
\Delta \Omega & = & \frac{1}{2}k_B T\sum_{i,j}
      \int_{0}^{d}{\rm d}z \int_{0}^{d}
        {\rm d}z'\, c_{i}(z)c_{j}(z')B_{ij}(z'-z)
\end{eqnarray}

\noindent We now imagine that the separation between the two
plates is increased from $d$ to $d+\delta z$ by adding a `slice'
of width $\delta z$ between the planes $z_{0}$ and $z_{0}+\delta
z$. We map the regions $0 \le z \le z_{0}$ and $z_{0} \le z \le d$
in the original system to the regions $0 \le z \le z_{0}$ and
$z_{0}+\delta z \le z \le d+\delta z$ in the modified system,
respectively. We then have:

\begin{eqnarray}
\delta \Omega_{\rm PB} & = & \frac{\varepsilon}{4\pi}\int_{0}^{d}
     {\rm d}z \,
     \left(\frac{{\rm d}\Psi}{{\rm d}z}\right)
    \delta \left(\frac{{\rm d}\Psi}{{\rm d}z}\right)
    + k_B T\int_{0}^{d}{\rm d}z \, \sum_{i}
      \delta c_{i} \ln \frac{c_{i}}{\zeta}
\nonumber \\
   & & + \delta z \left[
       \frac{\varepsilon}{8\pi}
       \left(\frac{{\rm d}\Psi}{{\rm d}z}\right)^{2} +
       k_B T \sum_{i} c_{i}\left(
          \ln \frac{c_{i}}{\zeta}-1 \right) \right]
       _{z = z_{0}}
\end{eqnarray}

The first term can be integrated by parts. With the
conventions described above, the boundary terms can be
written as follows:

\begin{eqnarray}
& \displaystyle{
     \Psi \delta \left.\frac{{\rm d}\Psi}{{\rm d}z}
            \right|_{0}^{z_{0}} +
\Psi \delta \left.\frac{{\rm d}\Psi}{{\rm d}z}
             \right|_{z_{0}}^{d} =
} \nonumber \\
& \displaystyle{
      = \Psi(z_{0})\left[
      \frac{{\rm d}\Psi_{\rm new}}{{\rm d}z}(z_{0})-
      \frac{{\rm d}\Psi}{{\rm d}z}(z_{0}) -
      \frac{{\rm d}\Psi_{\rm new}}{{\rm d}z}
             (z_{0}+\delta z) +
      \frac{{\rm d}\Psi}{{\rm d}z}(z_{0})
      \right]
} \nonumber \\
& \displaystyle{
      = -\Psi(z_{0})
      \frac{{\rm d}^{2} \Psi}{{\rm d}z^{2}}(z_{0})
      \delta z
}
\end{eqnarray}

\noindent where use of the boundary conditions at $z = 0$ and $z =
d$ has been made. Using this relation and the Poisson equation
(\ref{eq:Poisson}), we obtain:

\begin{eqnarray}
\delta \Omega_{\rm PB} & = & \delta z \left[
      \sum_{i}e_{i}\Psi c_{i} +
      \frac{\varepsilon}{8\pi}\left(
           \frac{{\rm d}\Psi}{{\rm d}z}\right)^{2}
      + k_B T \sum_{i}c_{i}\left(
          \ln \frac{c_{i}}{\zeta} - 1 \right)
      \right]_{z_{0}}
\nonumber \\
  & & + \int_{0}^{d}{\rm d}z\, \sum_{i}\delta c_{i}
   \left[ e_{i}\Psi + k_B T
     \ln \frac{c_{i}}{\zeta} \right]
\end{eqnarray}

\noindent
To compute $\delta \Delta \Omega$, $\Delta \Omega$ can be
separated to the following three terms:

\begin{eqnarray}
\Delta \Omega & = & \frac{1}{2}k_B T\sum_{i,j}
      \int_{0}^{z_{0}}{\rm d}z \int_{0}^{z_{0}}
        {\rm d}z'\, c_{i}(z)c_{j}(z')B_{ij}(z'-z)
\nonumber \\
  & &  +\frac{1}{2}k_B T\sum_{i,j}
      \int_{z_{0}}^{d}{\rm d}z \int_{z_{0}}^{d}
        {\rm d}z'\, c_{i}(z)c_{j}(z')B_{ij}(z'-z)
\nonumber \\
  & &  +k_B T\sum_{i,j}
      \int_{0}^{z_{0}}{\rm d}z \int_{z_{0}}^{d}
        {\rm d}z'\, c_{i}(z)c_{j}(z')B_{ij}(z'-z)
\label{eq:deltaf2}
\end{eqnarray}

\noindent The variation of $c_{i}$ in these three
terms gives:

\begin{equation}
\delta \Delta \Omega_1 =  k_B T\sum_{i,j}
      \int_{0}^{d}{\rm d}z \int_{0}^{d}
        {\rm d}z'\, c_{i}(z)\delta c_{j}(z')
        B_{ij}(z'-z)
\end{equation}

\noindent The variation of the third term in Eq.\
(\ref{eq:deltaf2}) gives two additional contributions, one from
the variation of $B(z'-z)$ under the insertion of the `slice' at
$z_{0}$:

\begin{equation}
\delta \Delta \Omega_2 = \delta z \cdot k_{B}T
     \sum_{i,j}
     \int_{0}^{z_{0}}{\rm d}z \int_{z_{0}}^{d}
     {\rm d}z'\, c_{i}(z)c_{j}(z')
     \frac{{\rm d}B_{ij}}{{\rm d}z}(z'-z)
\end{equation}

\noindent and the other from the integration over the `slice'
itself:
\begin{equation}
\delta \Delta \Omega_3 = \delta z \cdot k_{B}T
     \sum_{i,j}
     \int_{0}^{d}{\rm d}z \,
         c_{i}(z_{0})c_{j}(z)B_{ij}(z-z_{0})
\end{equation}

\noindent Summing up all the contributions to $\delta \Omega$
we have:



\begin{eqnarray}
\delta \Omega & = & \delta \Omega_{\rm PB} +
      \delta \Delta \Omega_1 +
      \delta \Delta \Omega_2 + \delta \Delta \Omega_3
\nonumber \\
   & = & \delta z \sum_{i}c_{i}(z_{0}) \times
       \left\{ \frac{\varepsilon}{8\pi}
       \left(\frac{{\rm d}\Psi}{{\rm d}z}\right)^{2}(z_{0})
       + e_{i}\Psi(z_{0})
       + k_B T\left[\ln
       \frac{c_{i}(z_{0})}{\zeta}-1\right]  
        + k_B T\sum_{j}\int_{0}^{d}dz\,
           c_{j}(z)B_{ij}(z-z_{0}) \right\}
\nonumber \\
 & & + \int_{0}^{d}dz\, \sum_{i}\delta c_{i}(z) \times
     \left\{ e_{i}\Psi(z) +k_B T \ln
        \frac{c_{i}(z)}{\zeta} 
       + k_B T\sum_{j}\int_{0}^{d}dz'\,
           c_{j}(z')B_{ij}(z'-z) \right\}
\nonumber \\
 & & + \delta z \, k_B T \sum_{ij}
    \int_{0}^{z_{0}}{\rm d}z \int_{z_{0}}^{d}{\rm d}z' \,
    c_{i}(z)c_{j}(z')\frac{{\rm d}B_{ij}}{{\rm d}z}(z'-z)
\end{eqnarray}

\noindent Using the equilibrium equation (\ref{eq:c1d})
this reduces to:

\begin{eqnarray}
-\frac{\delta \Omega}{\delta z} & = &
    k_{B}T\sum_{i}c_{i}(z_{0})
    - \left.\frac{\varepsilon}{8\pi}\left(\frac{{\rm d}\Psi}
      {{\rm d}z}\right)^2 \right|_{z_{0}}
\nonumber \\
  & & - k_{B}T\sum_{ij}\int_{0}^{z_{0}}{\rm d}z
                   \int_{z_{0}}^{d}{\rm d}z' \,
                   c_{i}(z)c_{j}(z')
                   \frac{{\rm d}B_{ij}}{{\rm d}z}(z'-z)
\end{eqnarray}

\noindent This result can be readily generalized to the case
of several ion species, as in Eq.~(\ref{eq:pressurein}).

\begin{multicols}{2}

%
%
%
%
%
%

\begin{thebibliography}{10}
\bibitem{Footnote1}
In Refs.~\onlinecite{Marcelja97Nature,Marcelja98,Marcelja97} there
was an error in the inclusion of the hydration pressure, making it
much smaller than its real value. The corrected data,
kindly provided to us by S. Mar{\u{c}}elja, is used in the 
comparison of the present work with the AHNC results.
\end{thebibliography}





\end{multicols}

\newpage

\setlength{\parskip}{\baselineskip}

\noindent {\bf Fig.~1}: Short-range effective potential between
${\rm Na}^{+}$ ion pairs, adapted
from Ref.~\onlinecite{Lyubartsev97}
using simulations in a bulk NaCl aqueous solution of concentration
$0.55$M, at room temperature \cite {Marcelja00pr}. The potential
is shown in units of $k_B T$, as a function of the distance between
the ion centers. The Coulomb interaction is subtracted to show
only the short-range hydration effect due to the water molecules.
For ion-ion separations below $2.9 \mbox{\AA}$ a hard core
interaction is taken.

\noindent {\bf Fig.~2}: Schematic description of the pair
potential model. An aqueous ionic solution confined between two
charged plates in (a) is replaced by ions in a continuum dielectric
medium with electrostatic and short-range interactions $u_{ij}(r)
= u_{ij}(|{\bf r}|)$ in (b). The coordinates $z = 0$ and $z = d$
designate the contact positions of the ions with the plates. The
distance of closest approach is equal to $d_{\rm hc}/2$, where
$d_{\rm hc}$ is the hard-core diameter of the ions.

\noindent{\bf Fig.~3}: The effective
layer-layer interaction $B(z)$ in a planar geometry,
as obtained
from the potential of Fig.~1
using Eq.~(\ref{eq:Bdefinition})
(solid line).
The oscillating structure of the radial potential
shown in Fig.~1 is apparent in the secondary minima
of $B(z)$.

\noindent {\bf Fig.~4}: Counterion density profile (solid line)
obtained from numerical solution of Eq.\ \ref{eq:c1d} with the
hydration interaction as of Fig.~3, plotted on a semi-log plot.
The bulk ion density is $c_b = 0.025$M and the surface charge is
$|\sigma| = 0.333 {\rm C/m^{2}}$. The dielectric constant is
$\varepsilon = 78$ and the temperature is $298$K. The distance
between the plates is $d = 50\mbox{\AA}$. The density profile is
symmetric about the mid-plane at $z = 25\,\mbox{\AA}$. The dotted
line shows the corresponding density profile obtained from the PB
equation. The symbols (`x') show the density profile obtained in
the AHNC approximation, using the same parameters.

\noindent {\bf Fig.~5}: The ratio of the
positive ion density obtained from
Eq.~(\ref{eq:c1d}) and the
value obtained from PB theory, for surface
charges $|\sigma| = 0.333{\rm C/m^{2}}$ (dashed line),
$0.1 {\rm C/m^{2}}$ (solid line) and
$0.0333 {\rm C/m^{2}}$ (dotted line). All other
parameters are as in Fig.~4.

\noindent {\bf Fig.~6}: The ratio between the positive ion density
obtained from Eq.\ (\ref{eq:c1d}) and its PB value, for plate
separations $d$ equal to (a) $50\,\mbox{\AA}$ (solid line),
$35\,\mbox{\AA}$ (dashed line),
$20\,\mbox{\AA}$ (solid line),
(b) $10\,\mbox{\AA}$ and (c) $5\,\mbox{\AA}$.
All other parameters are as in Fig.~4.
Each curve is shown between the plate at $z = 0$ and
the mid-plane $z = d/2$.

\noindent {\bf Fig.~7}: (a) Pressure between two plates with
surface charge $|\sigma| = 0.333 {\rm C/m^{2}}$, as a function of
the plate separation $d$, on a semi-log plot. All the parameters
are as in Fig.~4. The solid line shows the overall pressure $P$
obtained from Eq.\ (\ref{eq:pressured}). The dashed line shows the
contribution $P_{m}$ resulting from the mid-plane density and the
dotted line shows the PB pressure. (b) The same curves
on a linear scale, in the region where the overall pressure
becomes negative, {\it i.e.}, attractive.

\noindent {\bf Fig.~8}: The repulsive pressure between two
plates with surface charge $|\sigma| = 0.119 {\rm C/m^{2}}$, as a
function of the plate separation $d$. All other parameters are as
in Fig.~4. The solid line shows the overall pressure $P$, the
dashed line shows the contribution $P_m$ of the mid-plane density,
and the dotted line shows the results of PB theory.

\noindent {\bf Fig.~9}:
Comparison between the pressure obtained (a) in our model and
(b) in the AHNC approximation, using the same
short-range hydration potential (solid lines).
All the parameters are as in Fig.~7
($|\sigma| = 0.333 {\rm C/m^{2}}$).
The pressure is shown as a function
of the plate separation $d$. A semi-logarithmic
scale is used in the main plots and a linear scale
is used in the insets.
In (a) the dotted line shows the PB pressure. In (b) the
dotted line shows the pressure obtained in the AHNC
approximation when the ion-ion interaction includes only
the hard core and the electrostatic interactions.

\noindent {\bf Fig.~10}: Comparison between the 
pressure obtained in our model
(solid line) and the AHNC approximation
(dashed line),
for a surface charge $|\sigma| = 0.119 {\rm C/m^{2}}$.
All the parameters are as in Fig.~8.
The pressure is shown as a function of the inter-plate
separation $d$ using a semi-logarithmic plot. The dotted
line shows the PB pressure.

\noindent {\bf Fig.~11}: The hydration pressure $P_{\rm
hyd}$ as a function of the plate separation $d$ (solid line). All
the parameters are as in Fig.~4. The dashed line shows the
approximation to $P_{\rm hyd}$ obtained by replacing the ion
density in the integral of Eq.~(\ref{eq:phyd}) by the PB ion
density.

\noindent {\bf Fig.~12}: Repulsive pressure between two plates with surface
charge $|\sigma| = 0.25\,{\rm C/m^{2}}$, as a function of the
plate separation $d$, using a linear plot (solid line). All other
parameters are as in Fig.~4. The pressure is compared with the PB
pressure curve fitted to the large separation behavior, with
$|\sigma_{\rm eff}| \simeq 0.09\,{\rm C/m^{2}}$ (dashed line).

\setlength{\parskip}{0in}

\newpage

\begin{figure}[tbh]
\epsfxsize=0.5\linewidth \centerline{\hbox{
\epsffile{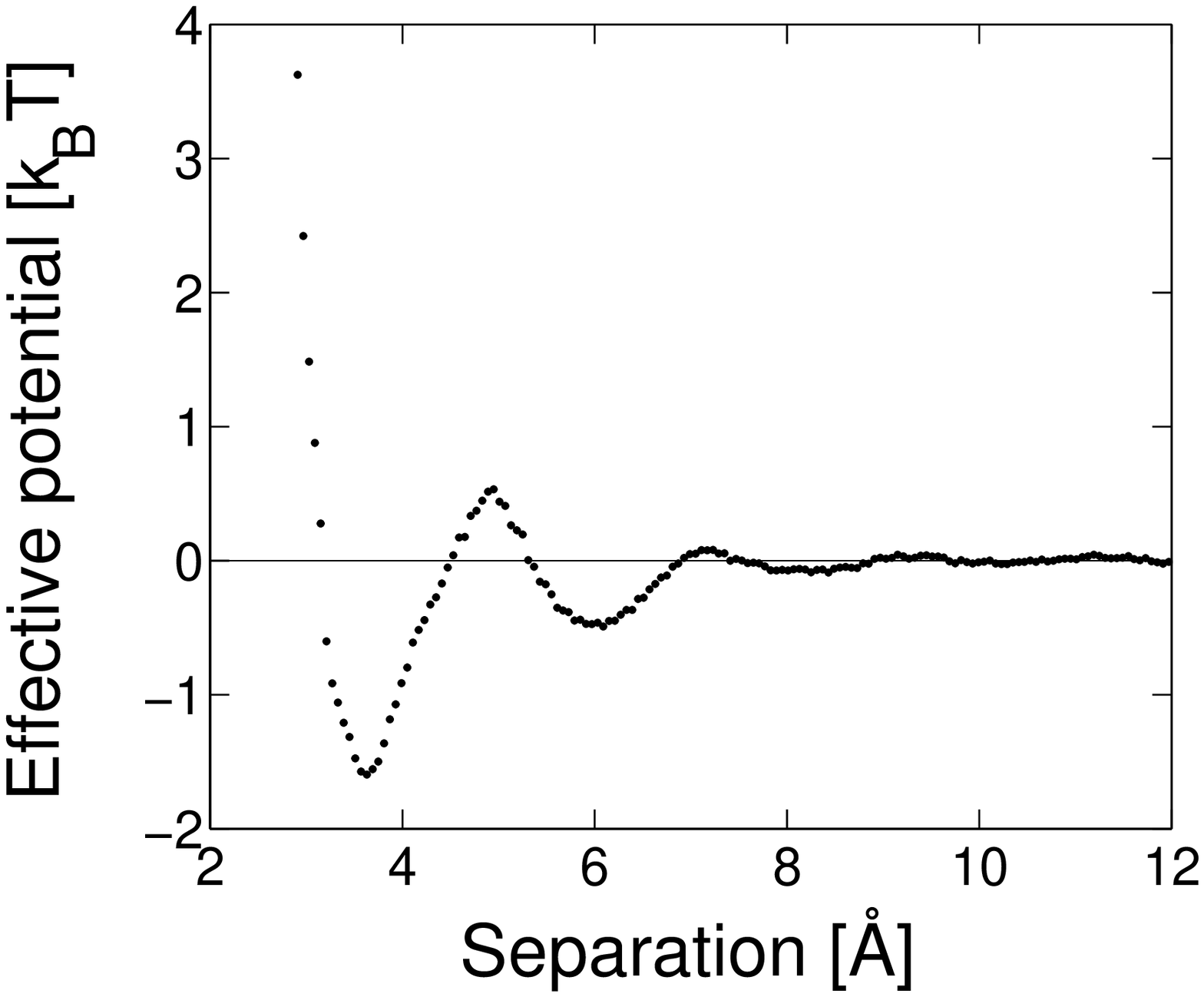} } }
\end{figure}
\vspace{0.45in}
\centerline{\large Fig.~1}

\vspace{0.45in}

\begin{figure}[tbh]
\epsfxsize=0.4\linewidth \centerline{\hbox{
\epsffile{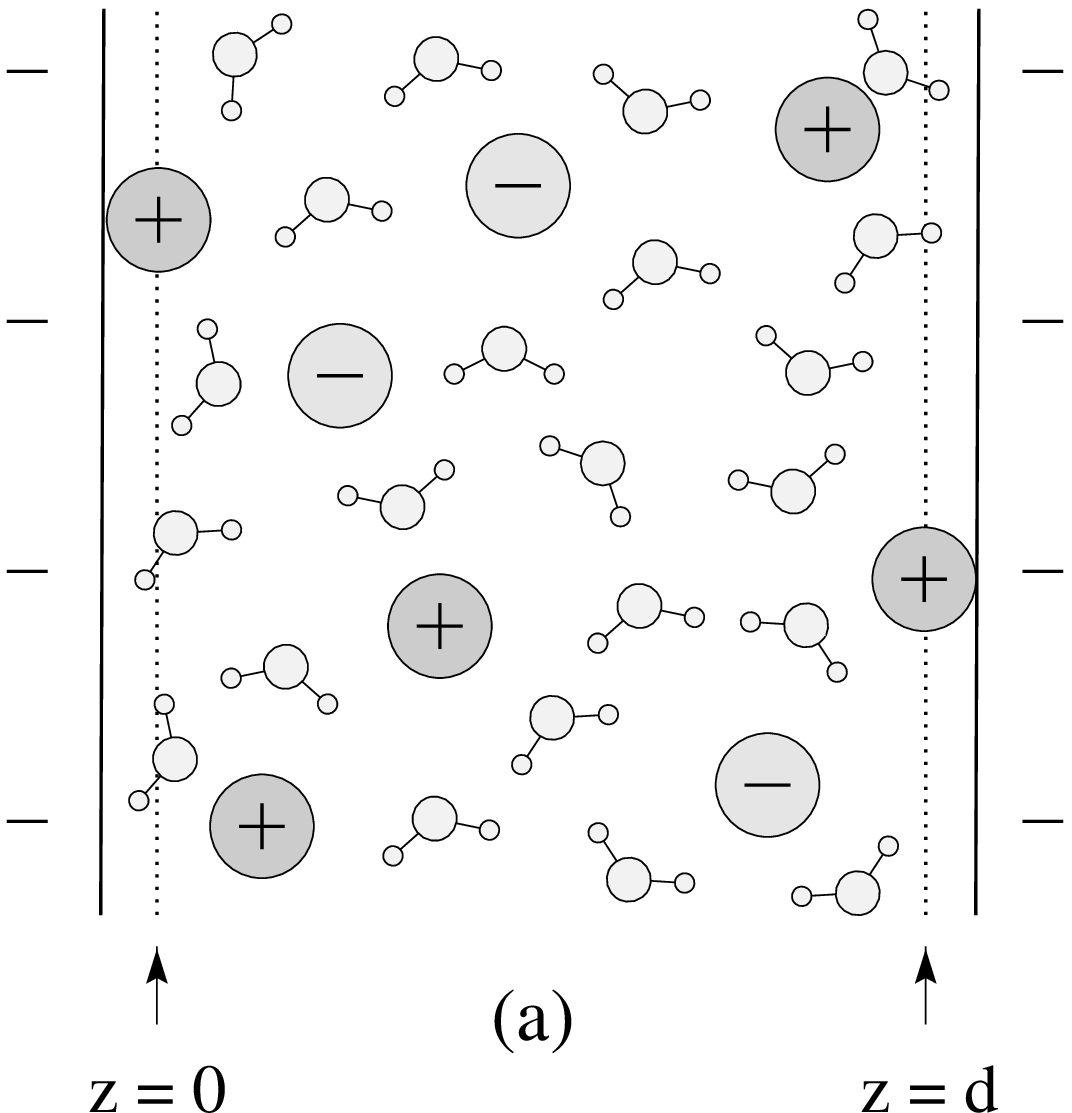} }
\hspace{0.45in}
\epsfxsize=0.4\linewidth
        \hbox{ \epsffile{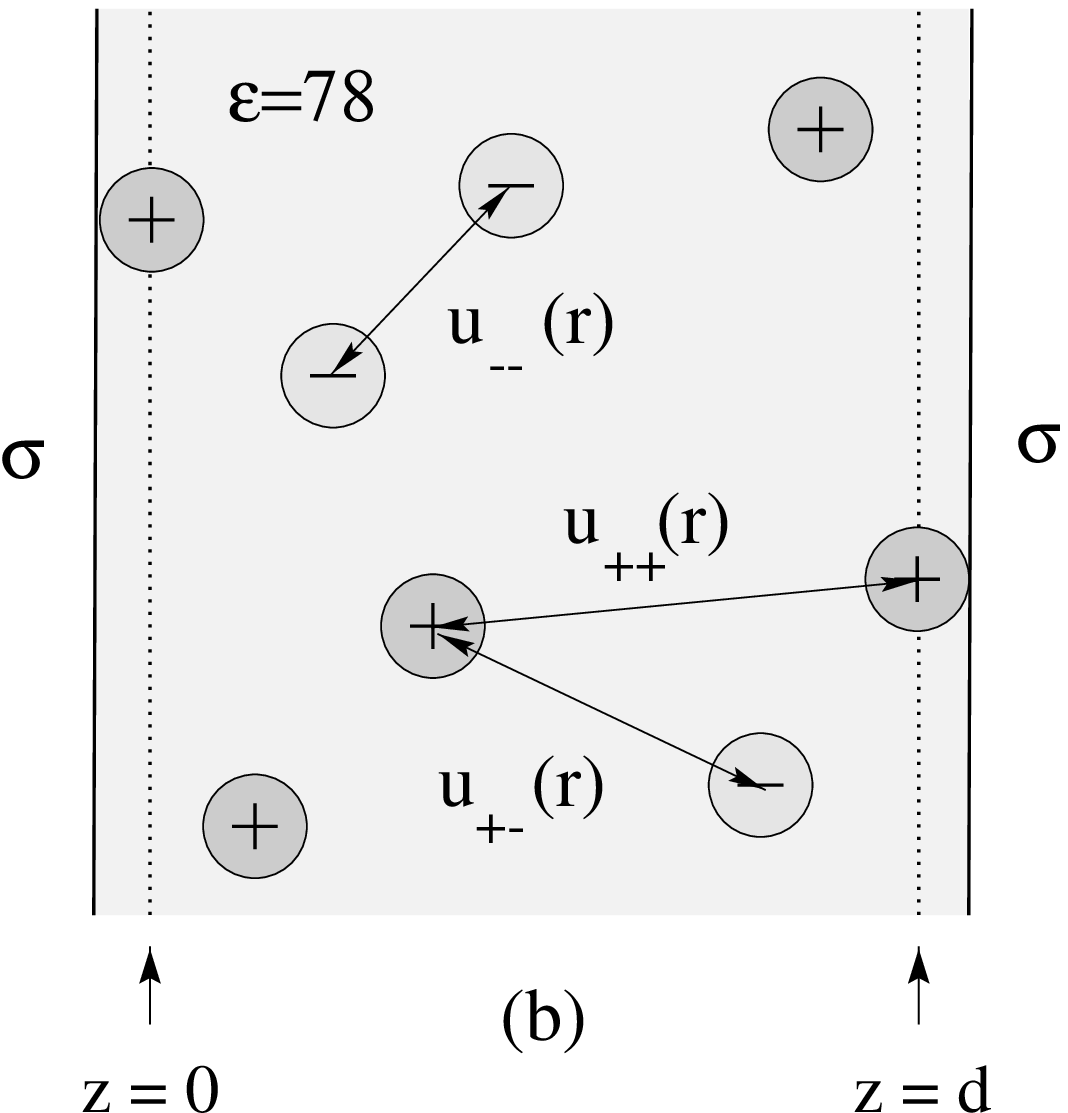} } }
\end{figure}
\vspace{0.45in}
\centerline{\large Fig.~2}

\vspace{0.45in} 
\centerline{\large Y.Burak + D.Andelman}

\newpage

\begin{figure}[tbh]
\epsfxsize=0.5\linewidth \centerline{\hbox{
\epsffile{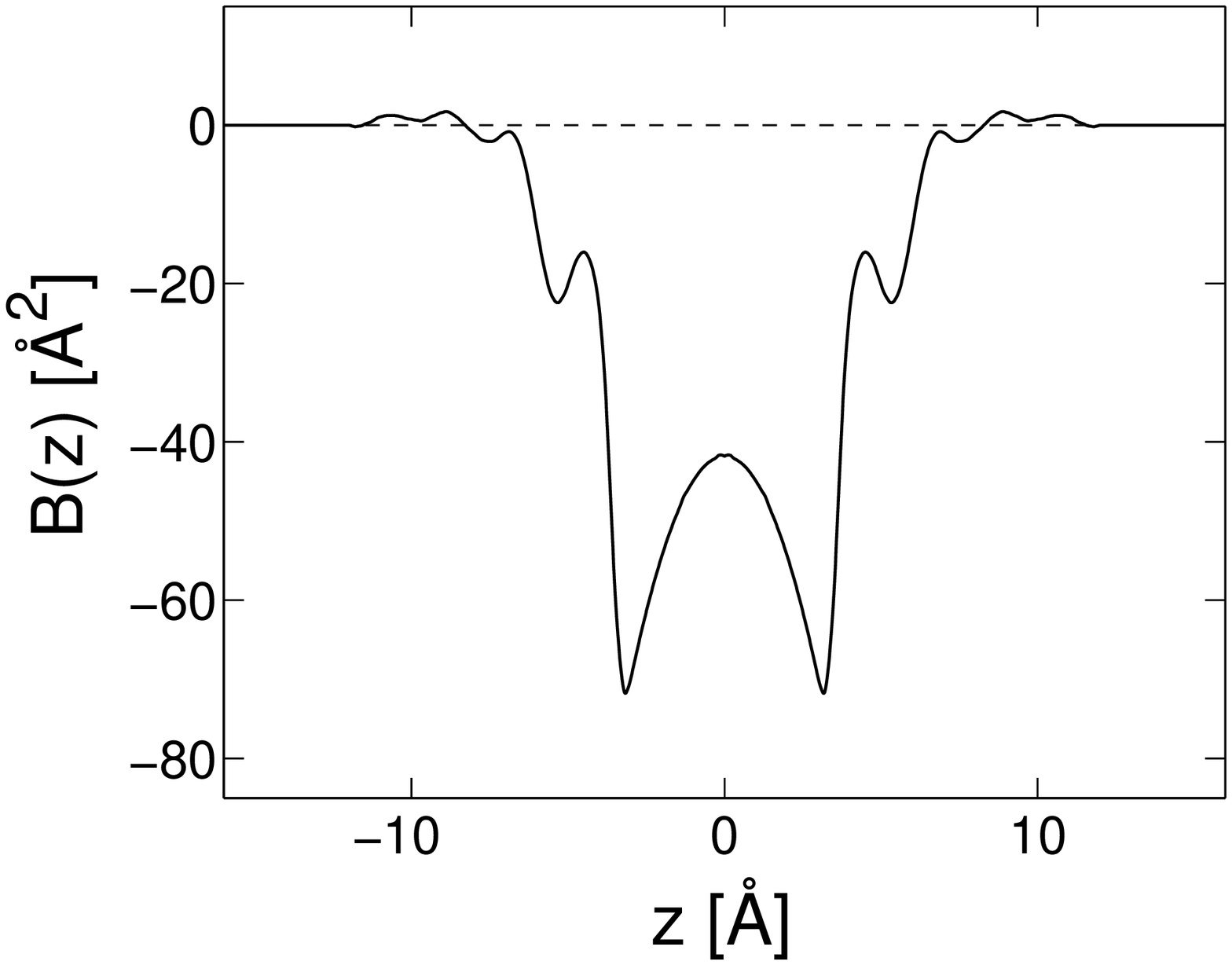} } }
\end{figure}
\vspace{0.45in}
\centerline{\large Fig.~3}

\vspace{1in}

\begin{figure}[tbh]
\epsfxsize=0.5\linewidth \centerline{\hbox{
\epsffile{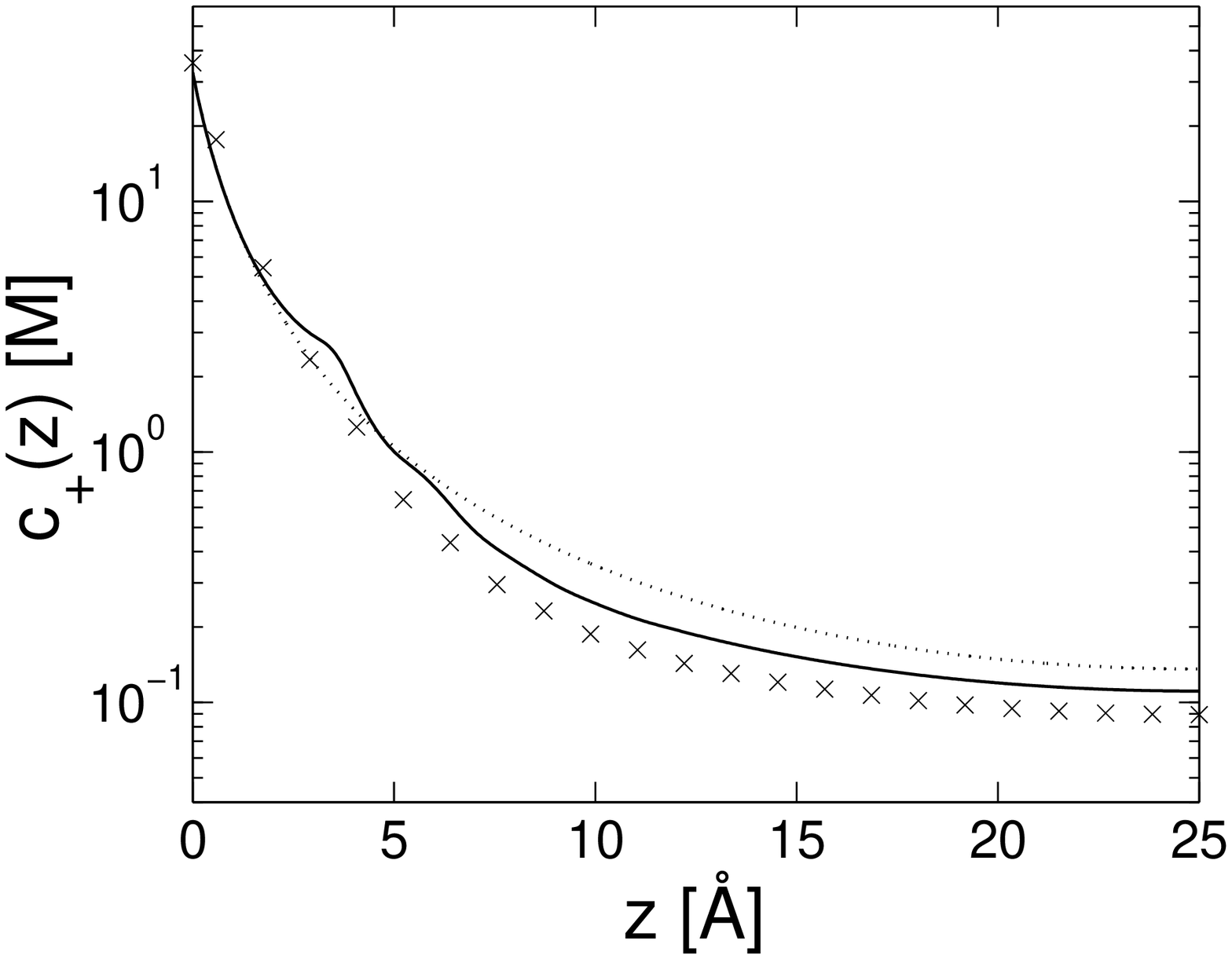} } }
\end{figure}
\vspace{0.45in}
\centerline{\large Fig.~4}

\vspace{0.45in}
\centerline{\large Y.Burak + D.Andelman}

\newpage

\begin{figure}[tbh]
\epsfxsize=0.5\linewidth \centerline{\hbox{
\epsffile{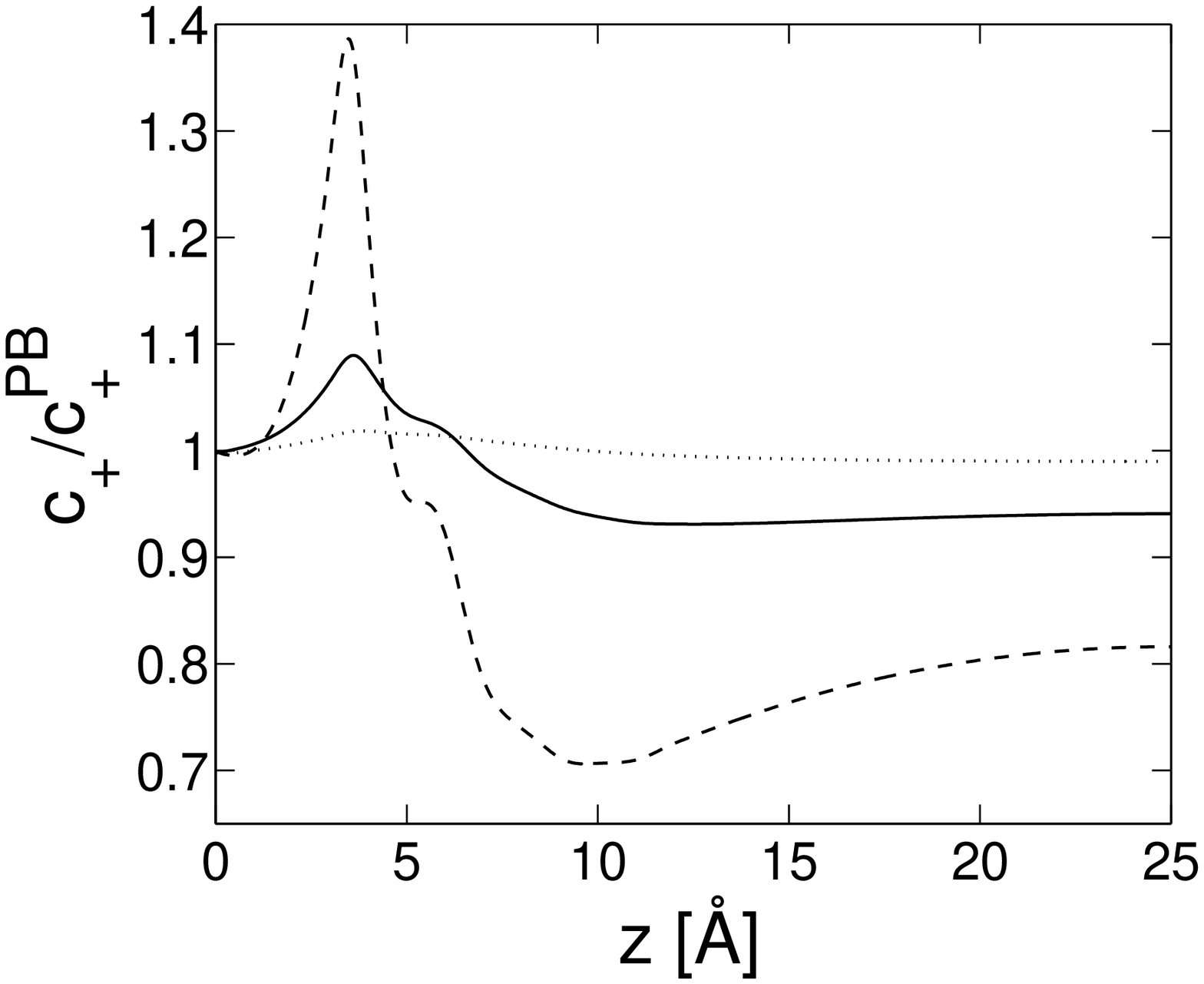} } }
\end{figure}
\vspace{0.45in}
\centerline{\large Fig.~5}

\vspace{1in}

\begin{figure}[tbh]
\epsfxsize=0.5\linewidth \centerline{\hbox{
\epsffile{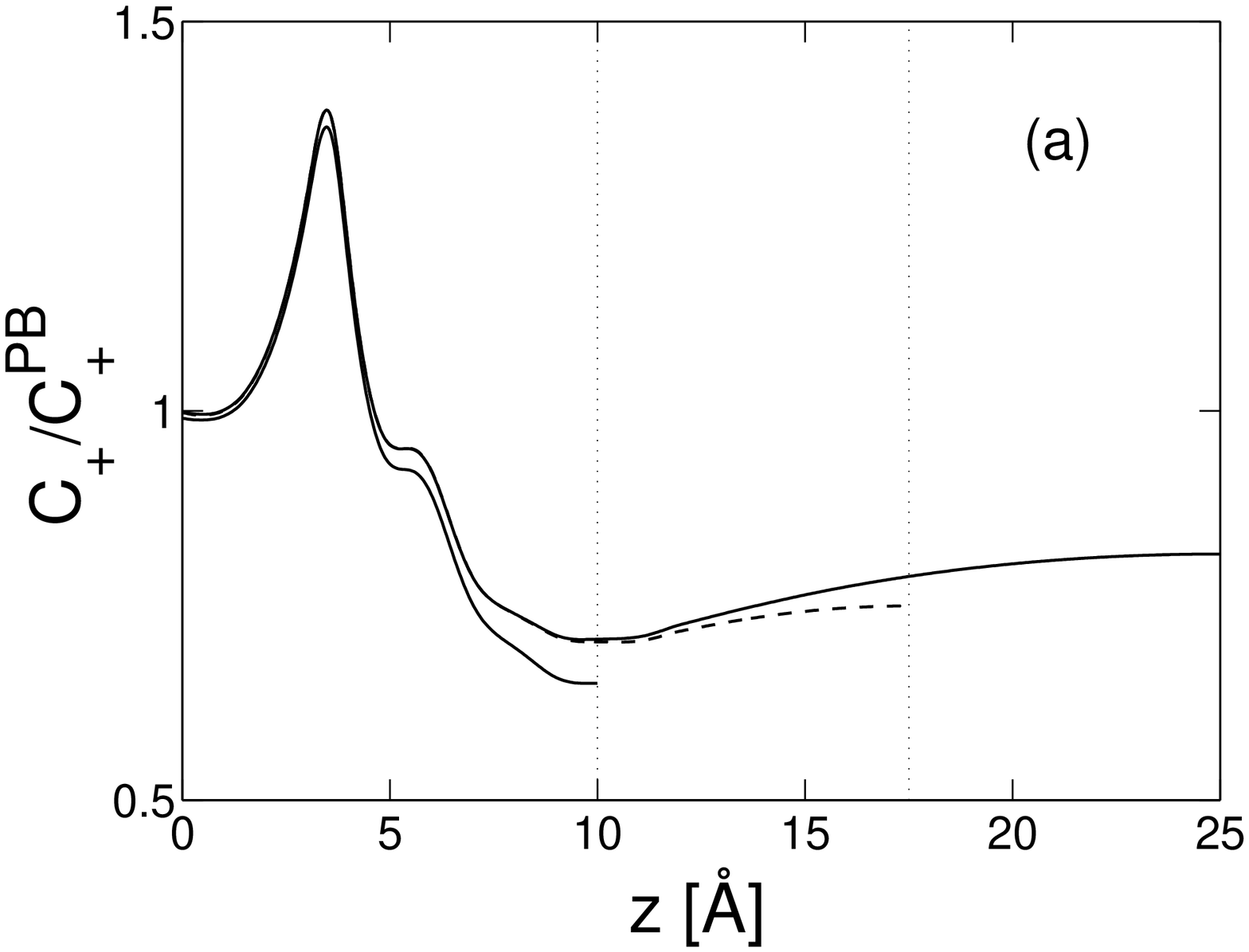} }
            \hspace{0.8cm}
            \epsfxsize=0.5\linewidth
            \hbox{ \epsffile{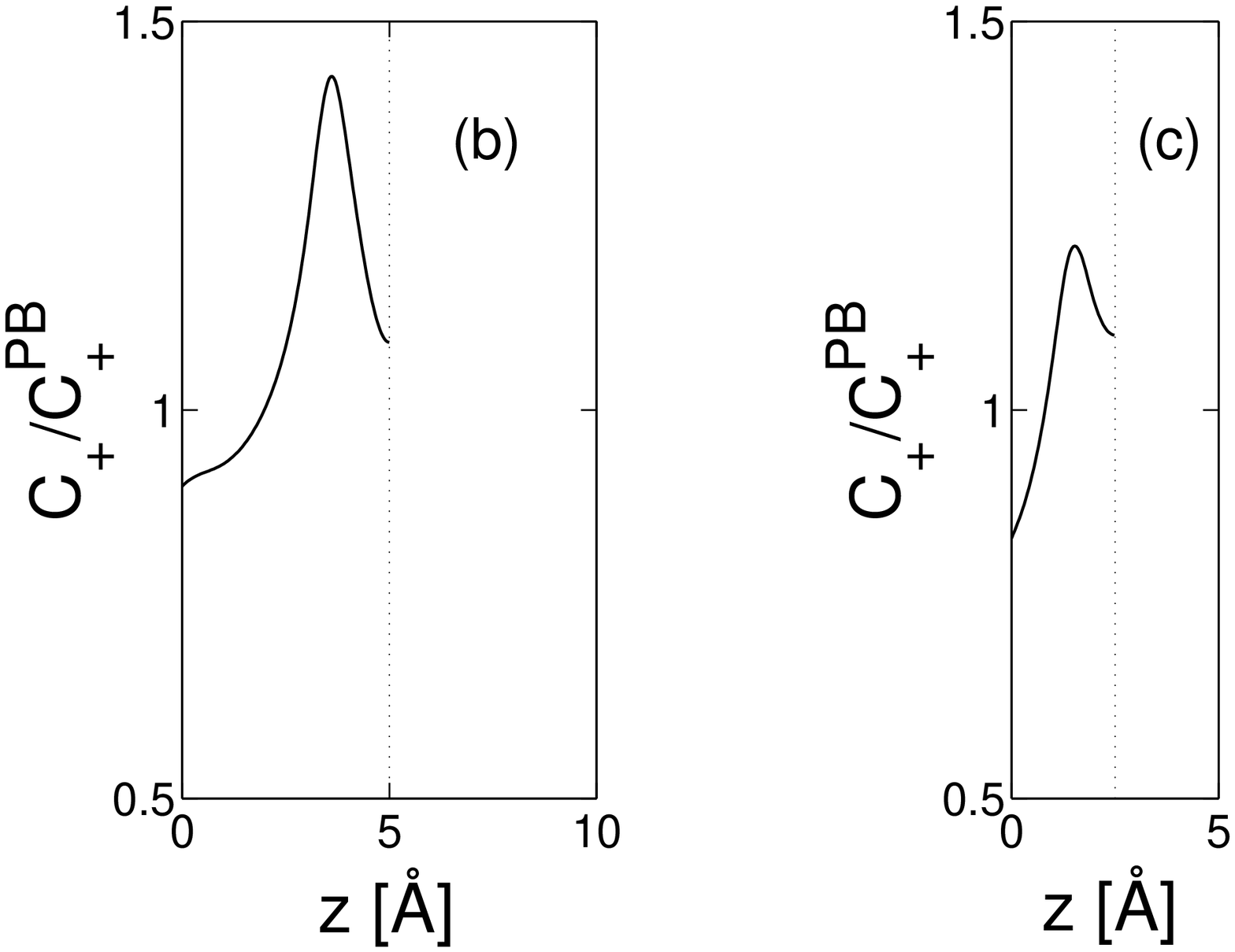} } }
\end{figure}
\vspace{0.45in}
\centerline{\large Fig.~6}

\vspace{0.45in}
\centerline{\large Y.Burak + D.Andelman}

\newpage

\begin{figure}[tbh]
\epsfxsize=0.5\linewidth \centerline{\hbox{
\epsffile{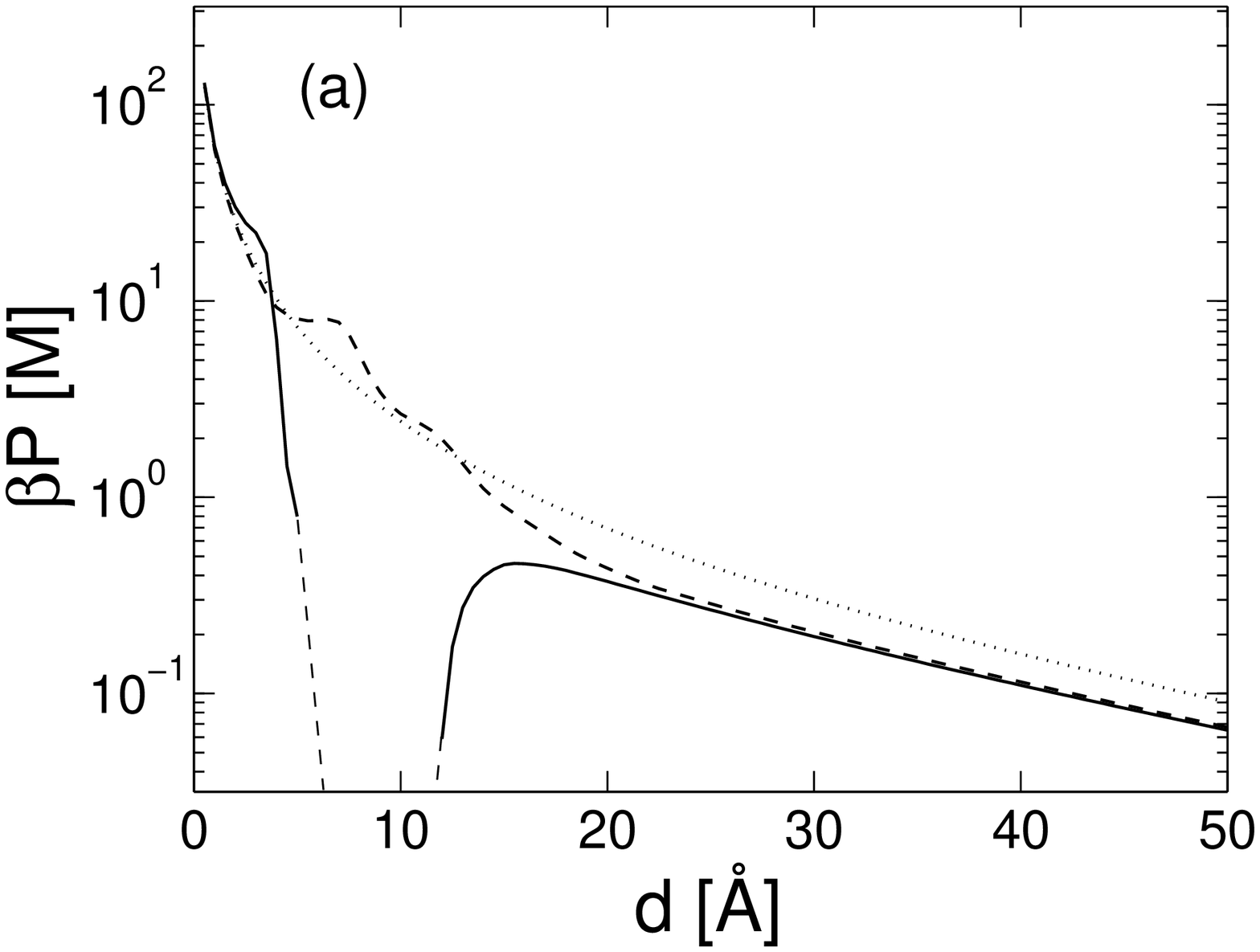} } \epsfxsize=0.5\linewidth
            \hbox{ \epsffile{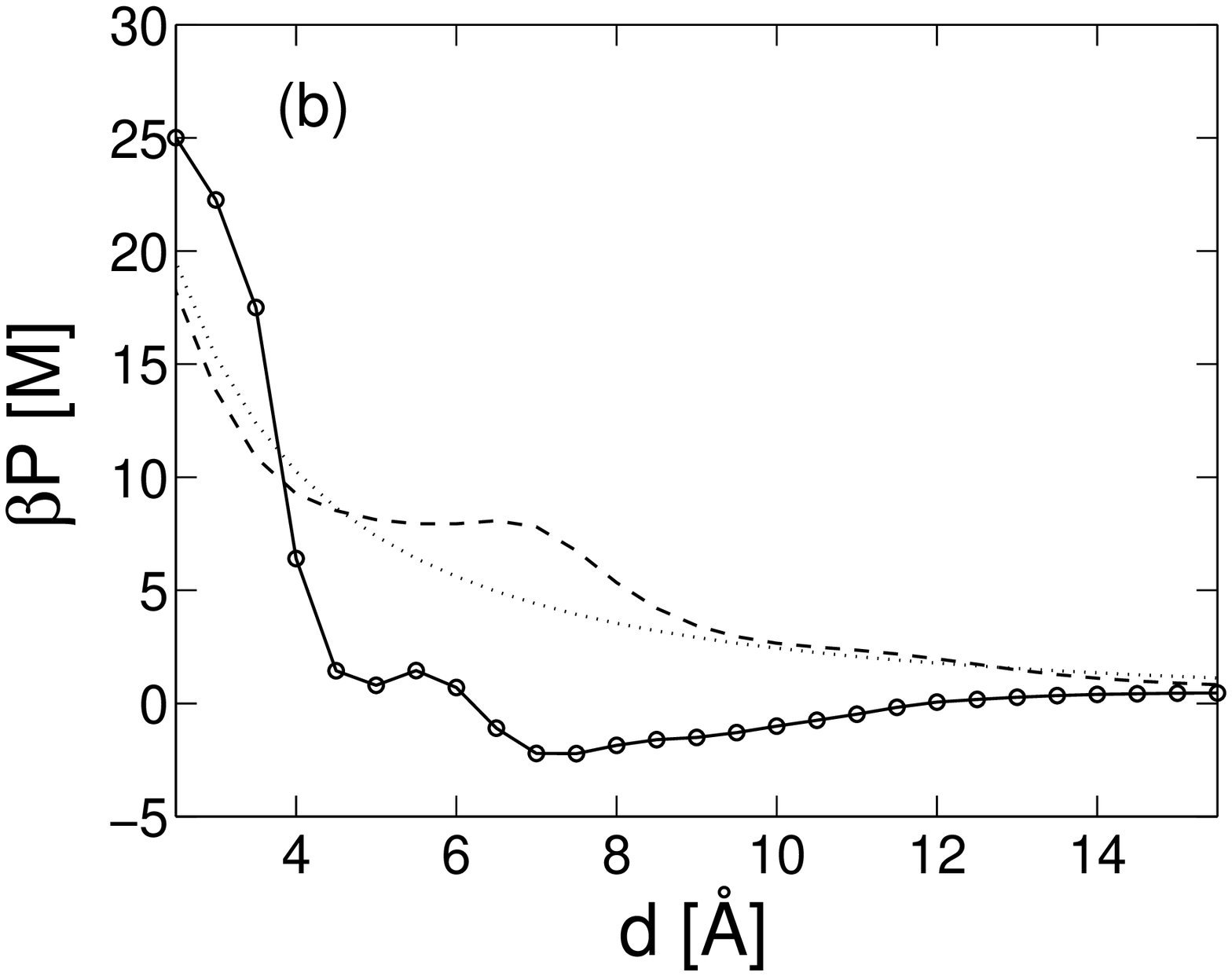} } }
\end{figure}
\vspace{0.45in}
\centerline{\large Fig.~7}

\vspace{1in}

\begin{figure}[tbh]
\epsfxsize=0.5\linewidth \centerline{\hbox{
\epsffile{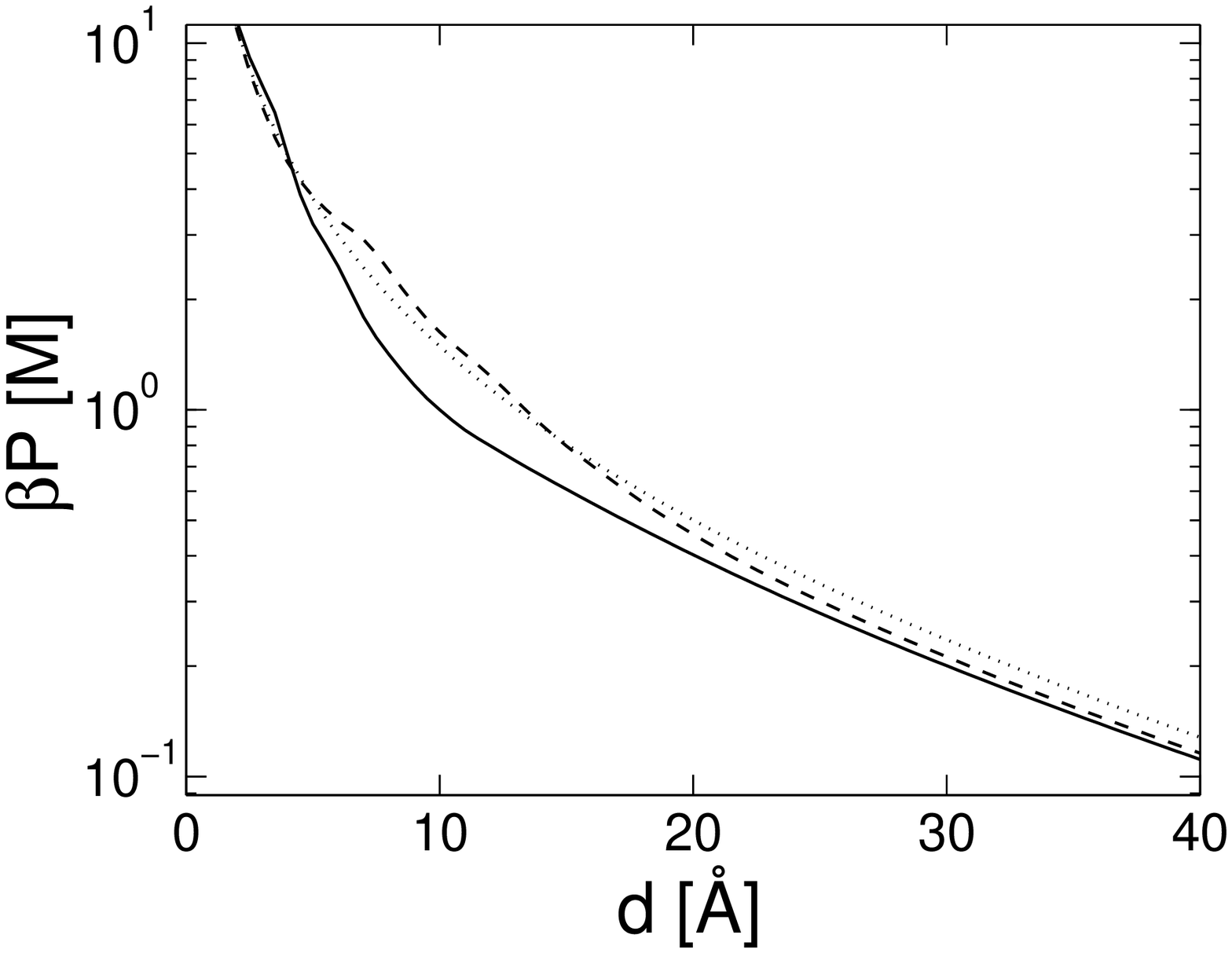} } }
\end{figure}
\vspace{0.45in}
\centerline{\large Fig.~8}

\vspace{0.45in}
\centerline{\large Y.Burak + D.Andelman}

\newpage

\begin{figure}[tbh]
\epsfxsize=0.5\linewidth \centerline{\hbox{
\epsffile{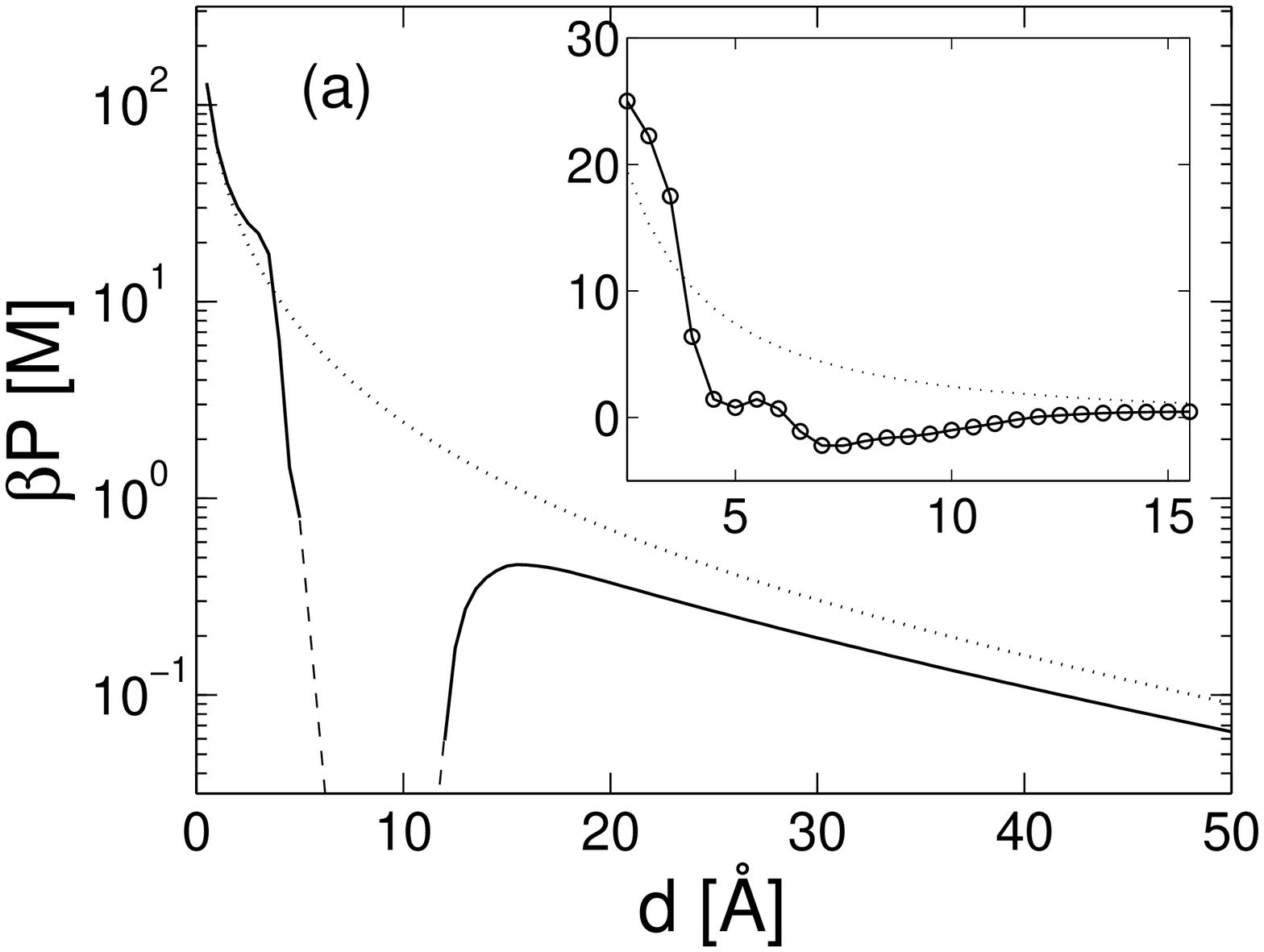} } \epsfxsize=0.5\linewidth
            \hbox{ \epsffile{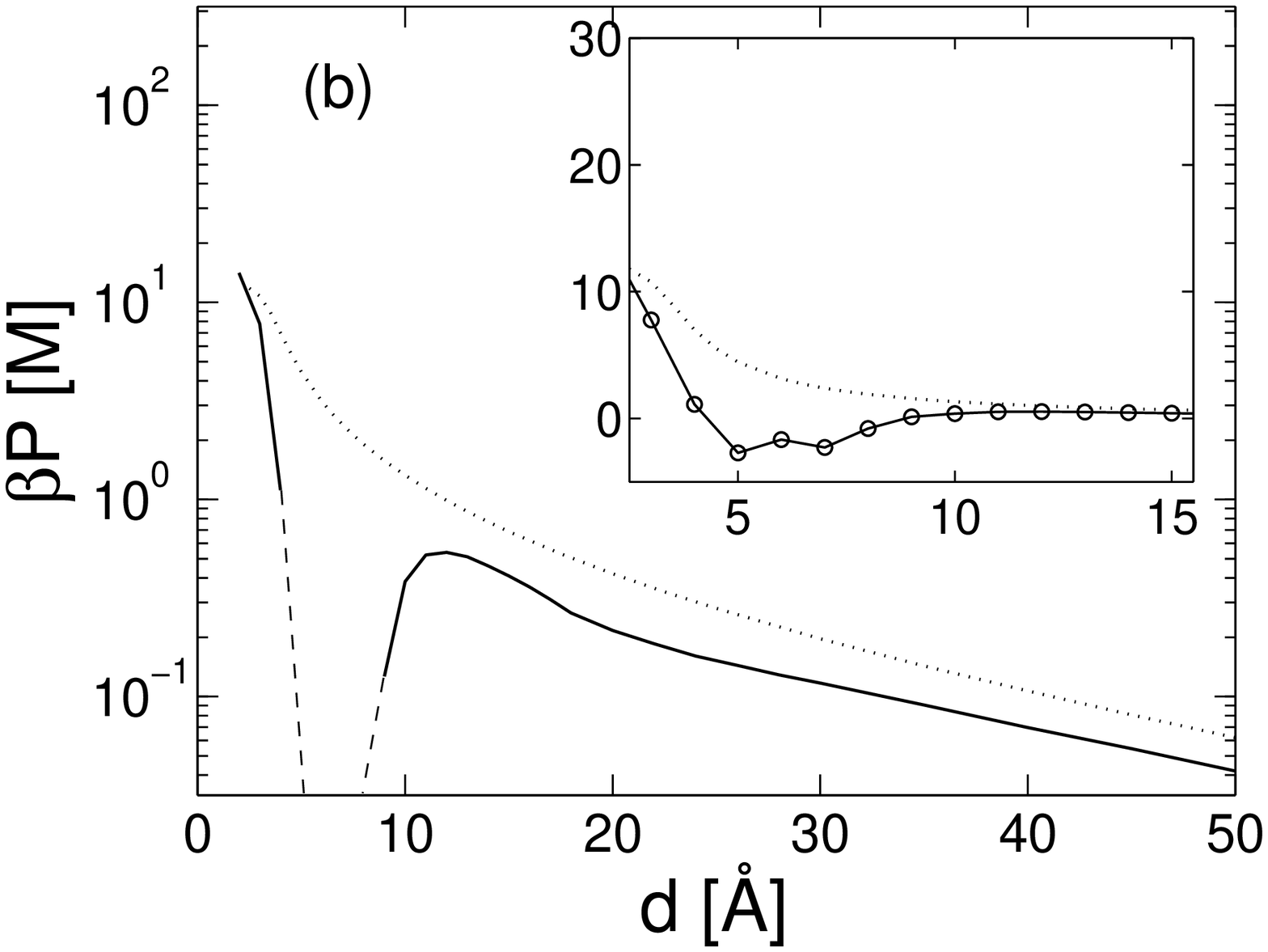} } }
\end{figure}
\vspace{0.45in}
\centerline{\large Fig.~9}

\vspace{1in}

\begin{figure}[tbh]
\epsfxsize=0.5\linewidth \centerline{\hbox{
\epsffile{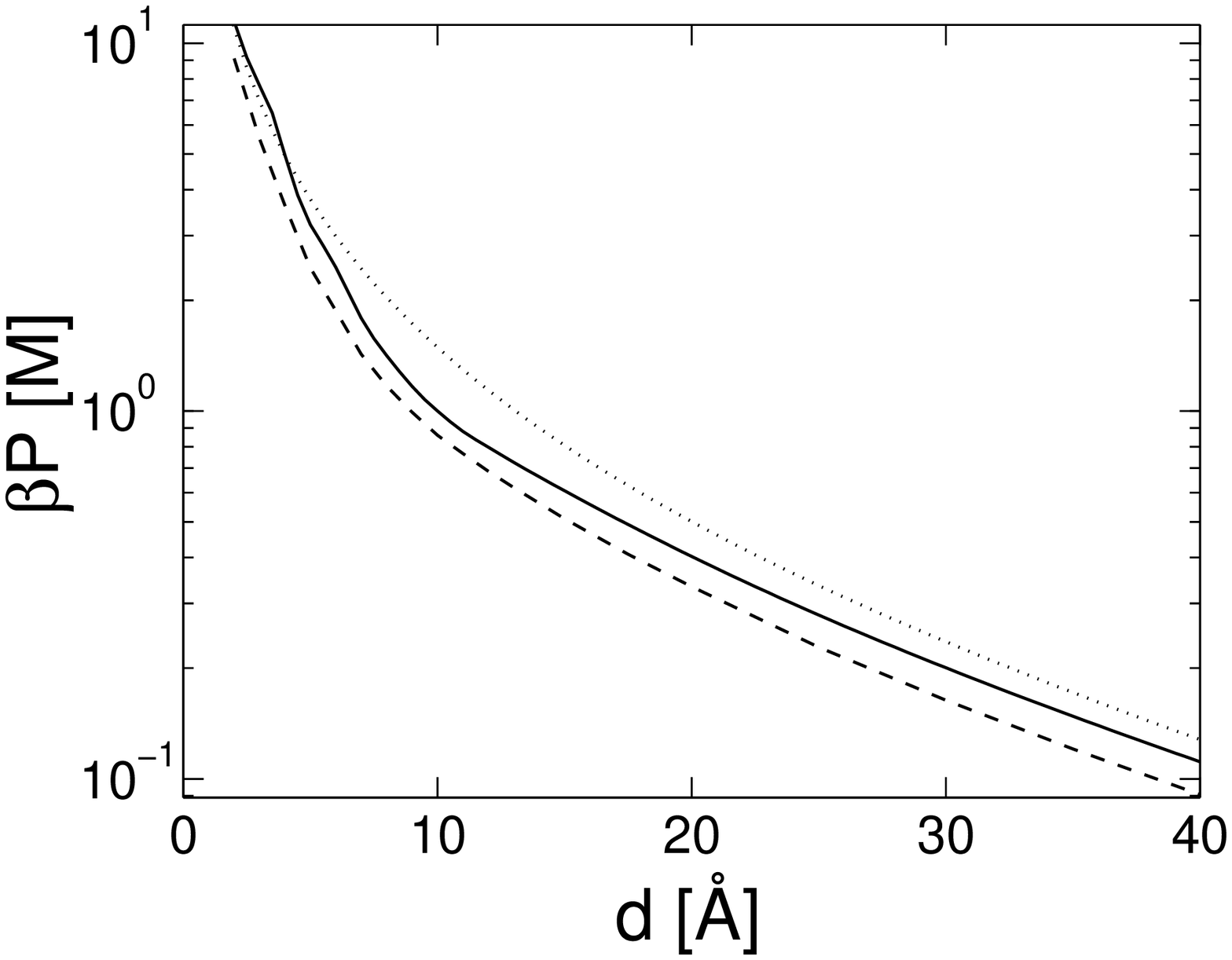} } }
\end{figure}
\vspace{0.45in}
\centerline{\large Fig.~10}

\vspace{0.45in}
\centerline{\large Y.Burak + D.Andelman}

\newpage

\begin{figure}[tbh]
\epsfxsize=0.5\linewidth \centerline{\hbox{
\epsffile{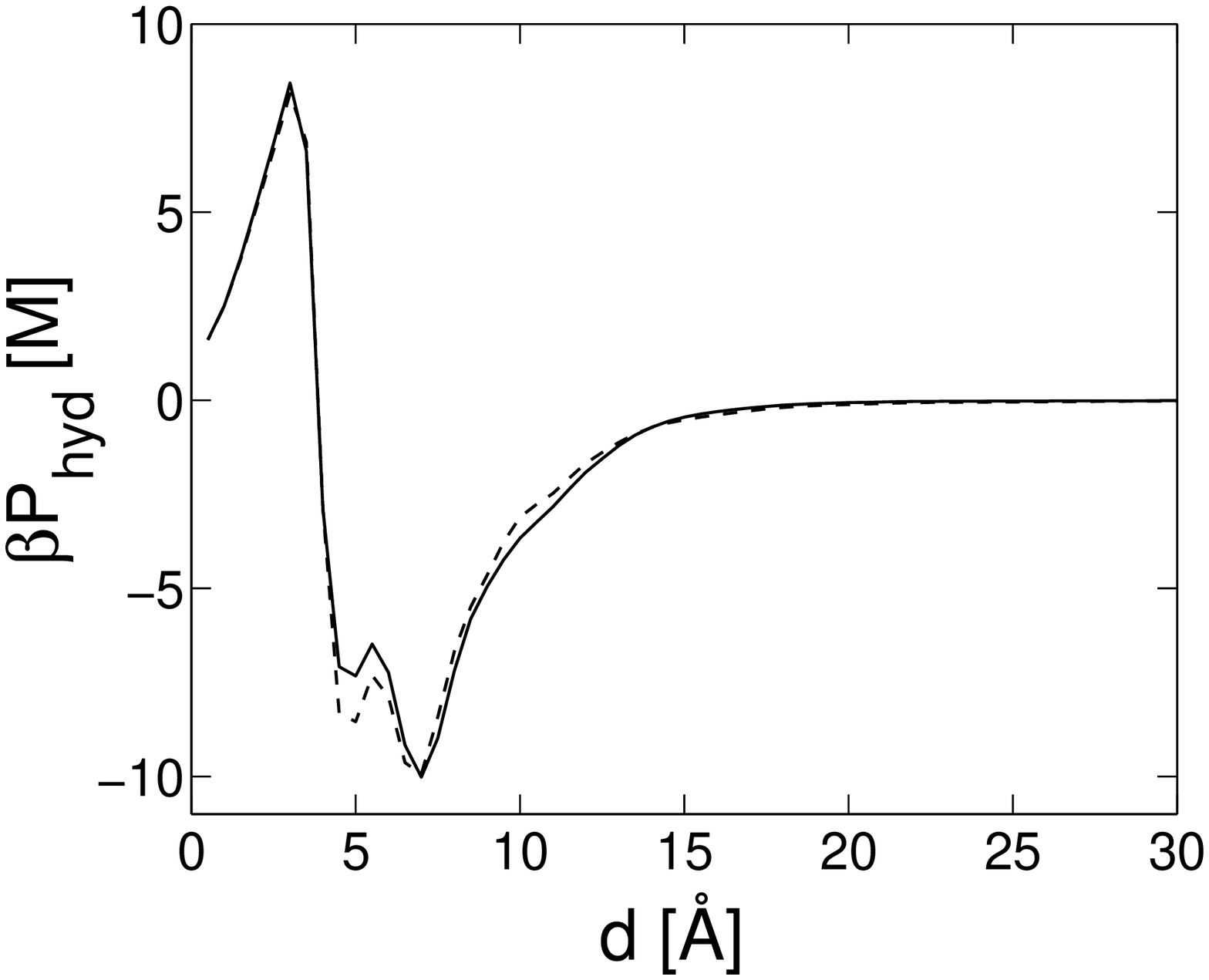} } }
\end{figure}
\vspace{0.45in}
\centerline{\large Fig.~11}

\vspace{1in}

\begin{figure}[tbh]
\epsfxsize=0.5\linewidth \centerline{\hbox{
\epsffile{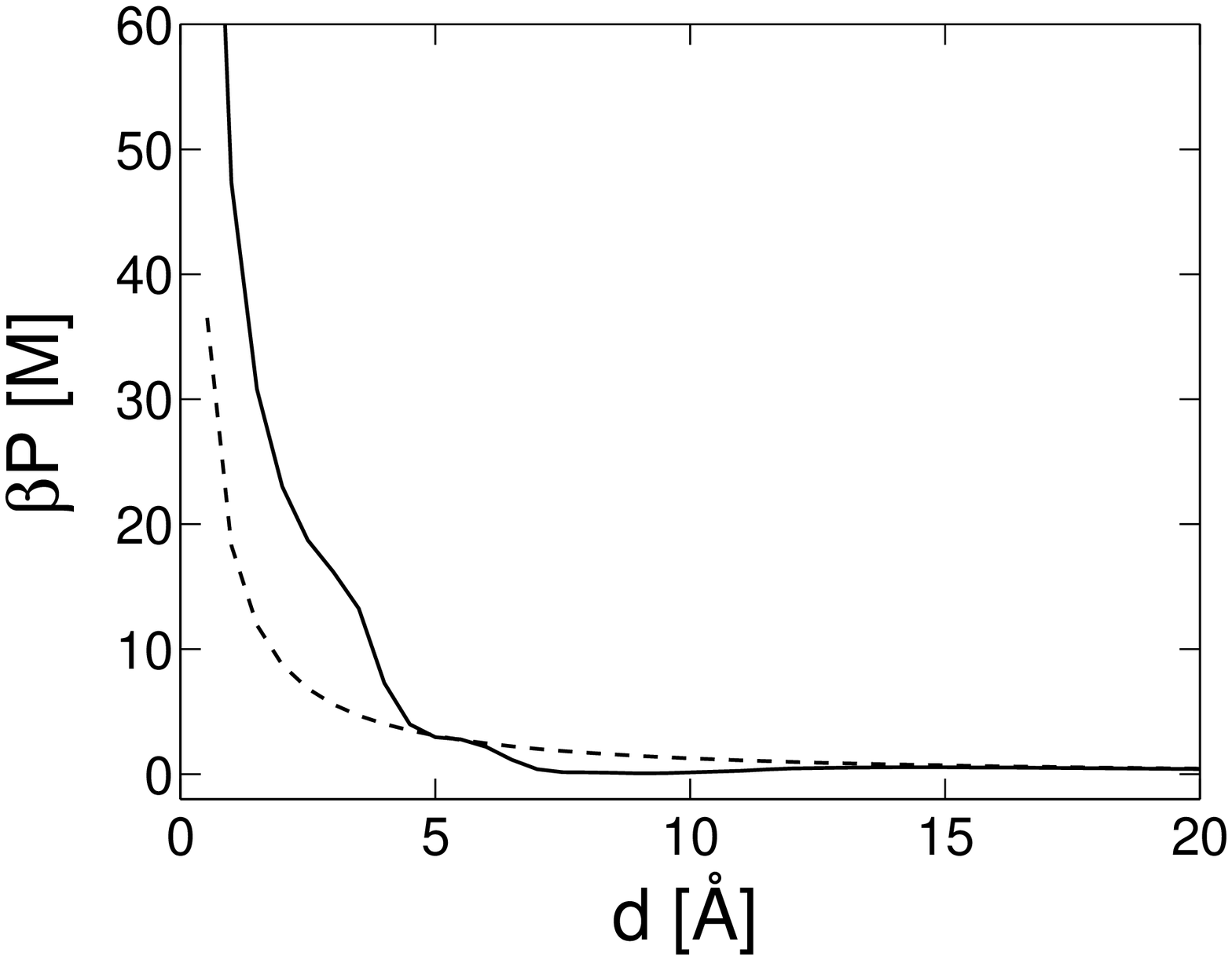} } }
\end{figure}
\vspace{0.45in}
\centerline{\large Fig.~12}

\vspace{0.45in}
\centerline{\large Y.Burak + D.Andelman}

\end{document}